\documentclass[twocolumn]{aastex62}

\usepackage{appendix}
\usepackage{amsmath}
\usepackage{subfigure}
\usepackage{amsfonts}
\usepackage{graphicx}
\usepackage{xcolor}
\usepackage{url}
\usepackage{color}

\newcommand{\beq}{\begin{equation}}
\newcommand{\eeq}{\end{equation}}
\newcommand{\bea}{\begin{eqnarray}}
\newcommand{\eea}{\end{eqnarray}}

\newcommand{\flash}{\texttt{FLASH}}

\newcommand{\msun}{\mathrm{M}_\odot}

\newcommand{\inv}{^{-1}}

\begin{document}

\title{Weak Shock Propagation with Accretion. III. \\ A Numerical Study On Shock Propagation \& Stability}

\author{Stephen Ro}
\affil{Astronomy Department and Theoretical Astrophysics Center, University of California, Berkeley, Berkeley, CA 94720}
\email{SR: stephensro@gmail.com}
\author[0000-0003-3765-6401]{Eric R.~Coughlin}
\altaffiliation{Einstein Fellow}
\affil{Columbia Astrophysics Laboratory, New York, NY, 80980}
\author{Eliot Quataert}
\affil{Astronomy Department and Theoretical Astrophysics Center, University of California, Berkeley, Berkeley, CA 94720}
\begin{abstract}
Core-collapse supernovae span a wide range of energies, from much less than to much greater than the binding energy of the progenitor star. As a result, the shock wave generated from a supernova explosion can have a wide range of Mach numbers. In this paper, we investigate the propagation of shocks with arbitrary initial strengths in polytropic stellar envelopes using a suite of spherically symmetric hydrodynamic simulations.   We interpret these results using the three known self-similar solutions for this problem:   the Sedov-Taylor blastwave describes an infinitely strong shock and the self-similar solutions from \cite{paper1} (Paper I) and \cite{paper2} (Paper II) describe a weak and infinitely weak shock (the latter being a rarefaction wave).   We find that shocks, no matter their initial strengths, evolve toward either the infinitely strong or infinitely weak self-similar solutions at sufficiently late times. For a given density profile, a single function characterizes the long-term evolution of a shock's radius and strength. However, shocks with strengths near the self-similar solution for a weak shock (from Paper I) evolve extremely slowly with time. Therefore, the self-similar solutions for infinitely strong and infinitely weak shocks are not likely to be realized in low-energy stellar explosions, which will instead retain memory of the shock strength initiated in the stellar interior.  

\end{abstract}
\keywords{hydrodynamics --- methods: analytical --- shock waves --- supernovae: general}

\section{Introduction}
\label{sec:introduction}
Self-similar solutions to systems of partial differential equations are valuable guides to complex physical problems. For instance, \cite{vonNeumann1941}, \cite{sedov1946} and \cite{1950RSPSA.201..159T} independently derived self-similar solutions for an energy-conserving point explosion in a cold homogeneous medium. These `Sedov-Taylor' blastwave solutions provide good descriptions for the early phase of a terrestrial explosion, and can also be used to describe the evolution of energetic supernovae \citep{1976ApJ...207..872C}.

While particularly energetic supernovae can be modeled by the Sedov-Taylor blastwave, there is growing evidence -- both observationally \citep{2011ApJ...738..154H} and theoretically \citep{O_Connor_2011, ertl16} -- that not all core-collapse events result in high-energy or even successful explosions. In these situations, the kinetic energy behind the blast can be comparable to or less than the binding energy of the star, and the Sedov-Taylor solutions cannot accurately reproduce the shock propagation or the evolution of the post-shock fluid. However, in the low-energy limit, there are other self-similar solutions that have been more recently discovered. For example, in supergiant stars, most of the stellar mass is concentrated in the core and the gravitational field in the extended, rarefied hydrogen envelope is approximately that of a point mass. The corresponding density profile of the envelope follows a simple power-law. For this configuration of a point-mass gravitational potential and a non-self-gravitating power-law density profile, \cite{paper1} -- hereon called `Paper I' -- derived self-similar solutions that describe the propagation of a weak shock (i.e., one with a Mach number that is only somewhat in excess of unity) that account for the binding energy of the envelope; these solutions also result in accretion onto the compact object at the origin. For convenience, we refer to these self-similar solutions as `CQR' and the self-similar solutions for strong shocks (e.g., Sedov-Taylor) as `SS'. In \cite{paper2} -- hereon called `Paper II' -- we derived another set of self-similar solutions that describe infinitely weak shocks. We refer to these as rarefaction wave (RW) solutions.

One specific physical scenario to which these self-similar solutions apply is a \emph{failed supernova}, in which the stalled, protoneutron star bounce shock (thought to be responsible for ejecting the envelope in a successful supernova explosion) is not revived. In this case, the formation of the protoneutron star still liberates $\sim few\times0.1\,\msun$ of mass in the form of neutrinos. Since the neutrinos escapes the star almost instantly, the over-pressurized envelope expands, and an acoustic pulse is launched from the inner regions of the star. This acoustic pulse steepens into a shock in the outer layers of the star \citep{Nadyozhin1980, 2013ApJ...769..109L, 2018MNRAS.476.2366F, 2018MNRAS.477.1225C}. 
If the Mach number of this secondary shock is very small, the ensuing hydrodynamic response is a rarefaction wave which merely informs the stellar envelope of the collapsed core. If the Mach number is only on the order of a few, then the shock can be adequately described by the CQR solution. By contrast, if the supernova is successful, the resulting shock typically has a very large Mach number, and the shock propagation will be well-described by the Sedov-Taylor blastwave.

In general, the shock from a supernova can have a seemingly arbitrary initial Mach number that may not map particularly well to any one of these self-similar regimes. We then ask how do these shocks evolve and what guidance, if any, do the self-similar solutions provide in their evolution? Our goal in this paper is to answer these questions with a suite of hydrodynamic simulations spanning a large range of explosion energies. We first define the physical problem and summarize the relevant self-similar solutions in \S\ref{sec:equations}. The numerical setup and parameters are in \S\ref{sec:numerics} and Appendix A. Results with discussion follow in \S\ref{sec:results1} and \S\ref{sec:results2} along with a summary in \S\ref{sec:conclusion}.

\section{Physical Problem and Solutions}
\label{sec:equations}
The derivations for the RW, CQR, and SS self-similar solutions are available in Papers I, II, \cite{DLBook1994}, \cite{ws93}, and various other sources. Here, we only describe the physical setup and present the relevant solutions. The notation here closely follows that of Paper II.

\subsection{Physical Setup}
We consider a spherically symmetric, non-self-gravitating, adiabatic, and motionless fluid with the following density and pressure structures:
\bea
\rho_1(r) &=& \rho_a \left(\frac{r}{r_a} \right)^{-n}, \label{eq:rho_ambient} \\
p_1(r) &=& p_a\left(\frac{r}{r_a} \right)^{-n - 1},
\label{eq:p_ambient}
\eea
where $\rho_a$ and $p_a$ are the density and pressure at radius $r=r_a$, and $n\ge0$ is the polytropic index. We take the adiabatic indices for the ambient fluid, $\gamma_1$, and post-shocked fluid, $\gamma_2$, to be identical and equal to $\gamma=1+n\inv$, implying that the gas is a pure polytrope (i.e., the adiabatic and polytropic indices are equal, which is the situation that is realized in the hydrogen envelopes of most supergiants). We note, though, that the results from Paper I and II do not require this choice. The ambient sound speed is then \beq
c_{1}(r) = \sqrt{\frac{\gamma p_1}{\rho_1} } = \sqrt{\frac{\gamma p_a r_a}{\rho_a r}}.
\label{eq:csound1}
\eeq
We also assume a point mass gravitational field or acceleration at all radii:
\beq
g = \frac{GM}{r^2},
\label{eq:gravity}
\eeq
where $M$ is the point mass and $G$ is Newton's constant. Substituting $GM=(n+1)p_a r_a/\rho_a$ renders the fluid motionless and relates the pressure of the ambient medium to the ambient density and the mass $M$.

\subsection{Self-Similar Solutions}
\subsubsection{Rarefaction Wave (RW) Solution}
\label{sec:rfwave}
In Paper II, we derive the self-similar solutions for a `shock' with zero strength and call this a rarefaction wave (RW). The RW solution is led by a sound wave of zero amplitude, or an acoustic node \citep{courant1999supersonic}, that propagates at the ambient sound speed, $c_1(r)$. After a RW arrives, the gas immediately falls inward and accretes onto the point mass at the origin. The RW solutions exist for any $n$.

\subsubsection{CQR Solution}
The shock-jump conditions \citep{Rankine01011870,ecole1887journal} demand that the velocity is positive immediately behind a shock expanding into a motionless fluid. If the shock is not strong, the shocked gas stagnates and eventually falls in toward the black hole. A sonic point forms where the infalling gas becomes supersonic with respect to the shock. The sonic point and shock-jump conditions provide two boundary conditions for the subsonic shocked flow which has a self-similar solution (Paper I). Analogous to standard spherical accretion \citep{1952MNRAS.112..195B}, the supersonic solution automatically connects to the black hole only if the sonic point conditions are satisfied correctly. 

The CQR solutions only exist between $2<n<3.5$. The shock expands at a velocity:
\beq
V_{\rm CQR}(R) = V_c\sqrt{\frac{GM}{R}},\label{eq:vshock_cqr}
\eeq
where $V_c$ is a dimensionless Eigenvalue that is unique for each $n$. As $n\rightarrow2$, $V_c$ diverges and the sonic point approaches the origin. This is consistent with a Sedov-Taylor blast wave in which the explosion is assumed to conserve a total energy (i.e., the shock and origin are in causal contact) that dwarfs the other energies in the problem. In the limit of $n=3.5$, $V_c\rightarrow(3.5)^{-1/2}$ and the Mach number,
\beq
M_{\rm CQR}(t) = \sqrt{n}V_c,
\label{eq:mach_cqr}
\eeq
approaches unity. In this limit, the CQR and RW solutions converge.

Integrating Eq.\,(\ref{eq:vshock_cqr}) yields the shock position for the CQR solution:
\beq
R_{\rm CQR}(t) = R_0\left(1+\frac{3}{2}\frac{V_{\rm c}\sqrt{GM}}{R_0^{3/2}}\left(t-t_0\right)\right)^{2/3},
\label{eq:rshock_cqr}
\eeq
given a reference position, $R_0$, at time, $t_0$. Since the self-similar solution is scale invariant, the temporal and spatial origin is arbitrary and, so, $t_0$ and $R_0$ can be set to zero and one, respectively, without loss of generality. The energy of the post-shock gas is not conserved for the CQR solution because the shock sweeps up gas with finite binding energy and accretion at the origin removes binding energy from the post-shock region.

\subsubsection{Strong Shock (SS) Solution}
\label{sec:sssss}
Self-similar solutions for energy-conserving, strong (i.e., Mach number much greater than one) spherical explosions have been described by numerous authors for $n<3$. The shock velocity and position follows
\bea
V_{\rm ST} &=& \frac{2 r_{\rm ST}}{(5 - n)t}, \label{eq:vshock_st}\\
R_{\rm ST}(t) &=& \left(\frac{E_0 t^2}{\beta \rho_a r_a^n} \right)^{\frac{1}{5 - n}},
\label{eq:rshock_st}
\eea 
for spherical geometry, where $E_0$ is the explosion energy, and $\beta$ is a unique value for given $\gamma$ and $n$. 

For $n>3$, the shock accelerates with a trailing sonic point. The total explosion energy behind the shock diminishes with time because the post-shock gas passes through the sonic point and loses causal contact with the shock. \cite{ws93} derive the self-similar solutions for strong shocks that satisfy the sonic point conditions. The temporal exponent for the shock position, $R\propto t^{\alpha}$, is always above unity (i.e., $\alpha_{\rm WS}>1$) and must be found numerically. On the other hand for $n<2$, strong shocks in a medium satisfying Eq.\,(\ref{eq:csound1}) (i.e., $c_1(r)\propto r^{-1/2}$) must weaken and eventually depart from the SS solution.

The post-shock gas structure depends on the polytropic and adiabatic indices\footnote{A gallery can be found on Frank Timmes' website: \url{http://cococubed.asu.edu/research\_pages/sedov.shtml}}. Material is distributed at all radii behind the shock for $n<n_h=(7-\gamma)/(\gamma+1)$. For $n\ge n_h$, the self-similar solutions have a vacuum interior below a contact discontinuity (CD) that resides at some dimensionless radius, $\xi_{\rm CD}=r/R$ \citep{1990ApJ...358..214G}. For $n> n_c=6/(\gamma+1)$, the density at the CD becomes infinite. In our setup where $\gamma=1+n\inv$, we have $n_h\simeq2.28$ and $n_c=2.5$. 

For a certain range of polytropic indices $3<n<n_g(\gamma)$, \cite{2010ApJ...723...10K} state that the solutions from \cite{ws93} do not exist (e.g., $n_g(\gamma\!=\!4/3)\!\simeq\!3.13$, $n_g(\gamma\!=\!5/3)\!\simeq\!3.26$). Instead, there is another class of self-similar solutions where the sonic point does not manifest and the shock does not accelerate (i.e., $R\propto t^1$). As the authors state, the hydrodynamic simulations converge very slowly for this setup. In the interest of our broader goals, we do not include polytropic indices in this range.

\subsection{Shock Trajectories}
\label{sec:trajectory}
A shock propagates along a space-time trajectory, $R(t)$, with velocity $V=dR/dt$. Re-writing this as the instantaneous temporal power-law of the shock position,
\beq
\alpha\equiv\frac{d{\rm log}(R)}{d{\rm log}(t)} = \frac{t}{R}V,
\label{eq:alpha}
\eeq
the shock trajectory grows as $R\propto t^\alpha$ assuming $\alpha$ varies slowly with time. Since the ambient sound speed follows $c_1\propto R^{-1/2}\propto t^{-\alpha/2}$ at the shock position, the Mach number of the shock trajectory evolves as
\beq
M_{\rm s} = \frac{V}{c_1} \propto t^{\frac{3}{2}\alpha -1}.
\label{eq:machtrajectory}
\eeq
We use Eq.\,(\ref{eq:machtrajectory}) to measure the growth of a shock's strength. Shocks that have $\alpha>2/3$ strengthen (i.e., increase in Mach number) with time, while shocks satisfying $\alpha<2/3$ weaken.  

SS solutions have $\alpha_{\rm ST}=2/(5-n)<1$ to conserve total energy ($n<3$) and $\alpha_{\rm WS}>1$ due to constraints imposed by the sonic point ($n>3$). We refer to these values as $\alpha_{\rm SS}$ unless the polytropic index is specified. Suppose strong shocks have temporal power-laws near $\alpha_{\rm SS}$. Since $\alpha_{\rm ST}<2/3$ for $n<2$, strong shocks decay in strength and will not resemble the Sedov-Taylor solution at late times. For $n>2$, strong shocks continue to strengthen and are expected to resemble the SS solution at late times.

A trajectory with finite and constant Mach number must have a constant temporal power-law of $\alpha=2/3$. Therefore, RW and CQR solutions have trajectories that expand as
\beq
R\propto t^{2/3}.
\label{eq:r_selfsimilar}
\eeq
Eq.\,(\ref{eq:machtrajectory}) is then not useful in measuring the growth of a shock with Mach numbers near the RW and CQR values. In Paper II, we present a detailed linear radial perturbation analysis for the CQR solution as well as the RW and SS solutions. The perturbed CQR solutions can be written as a linear sum of the unperturbed CQR solution and an infinite set of non-standard radial Eigenmodes. All of the Eigenmodes are found to rapidly decay with time except for the first and only growing Eigenmode. This is true for every $n$ where the CQR solution exists. Thus, the CQR solution is always linearly unstable to perturbations. We also find the RW solution is linearly unstable to perturbations for $n>3.5$ but stable otherwise.

A self-similar solution that is linearly unstable is unlikely to be realized at late times; this leads us to question its fate. Do such shocks evolve to another known self-similar solution or to a different solution entirely (self-similar or not)? In Paper II, we find that the Sedov-Taylor solution is linearly stable to radial perturbations. Yet, Eq.\,(\ref{eq:machtrajectory}) suggests that an initially strong shock must weaken for $n<2$ and, therefore, cannot resemble the Sedov-Taylor solution at late times. These points suggest that the linear stability of a self-similar solution does not necessarily determine its fate. In what follows, we employ hydrodynamic simulations to understand the long-term evolution of shocks that are not exactly self-similar.

\section{Numerical Setup}
\label{sec:numerics}
We employ the hydrodynamic code \flash\ (\citealt{2000ApJS..131..273F}) and the HLLC Riemann solver for our investigation. The simulation domain is a one-dimensional, spherical grid with uniform grid resolution. We assume an ideal equation of state where $\gamma = 1+n\inv$, $\gamma$ being the adiabatic index of the gas and $n$ the power-law index of the density of the ambient medium (i.e., $\rho \propto r^{-n}$). The initial density and pressure across the domain are defined by Eq.\,(\ref{eq:rho_ambient}) and (\ref{eq:p_ambient}). We use a constant point mass gravitational field as defined by Eq.\,(\ref{eq:gravity}). The inner boundary is at $r_{\rm min}/r_a=1$ with the Bondi-outflow condition described by \cite{2007ApJ...667..626K}. The fluid variables at the inner ghost zones are linearly extrapolated from the values at the inner boundary. We set the outer boundary at $r_{\rm max}/r_a=10^3$ and prescribe a reflecting boundary condition to maintain hydrostatic equilibrium. The initial pressure between $1\le r/r_a\le 1.5$ is set to a constant $p=(1+\delta p)\times p_a$, which introduces an over-pressured edge at $r/r_a=1.5$. This setup triggers a shock to form immediately and near the inner boundary. To simplify the notation, throughout the remainder of the paper we set $\rho_a=p_a=r_a=1$.

We run a suite of simulations which span the only two physical parameters in this problem: $n$ and $\delta p$. A table of simulation parameters is available at the end of the paper along with a brief discussion in Appendix A. The table includes values of the shock Mach number at $t=100$, which is a somewhat more intuitive way of gauging the shock strength than the value of $\delta p$. The `high' resolution simulations from Paper II (see their Table 3) are also included here. For large $n$ and small $\delta p$ we increase the resolution to better resolve the weak shock. The simulations are complete once $t=10^3$ or when the (strong) shock reaches the outer boundary. 

A true discontinuity cannot manifest on a discretized grid.  
We adopt the grid cell with the largest outward velocity as the shock location, $R$, which is always in close proximity to the cells with highest compression gradient. To quantify the Mach number of the shock, we take the pressure at the cell with maximum velocity as the immediate post-shock pressure, $p_2=p(R)$. The ambient pressure, $p_1$, is given by Eq.\,(\ref{eq:p_ambient}) using the shock location. With the shock jump condition for momentum conservation,
\beq
\frac{p_2}{p_1} = \frac{2\gamma M_{\rm s}^2 - (\gamma-1)}{\gamma+1},
\label{eq:shockjump}
\eeq
we can compute the instantaneous Mach number, $M_{\rm s}$, and, with Eq.\,(\ref{eq:csound1}), the velocity of the shock.

The temporal power-law of a shock's trajectory, Eq.\,(\ref{eq:alpha}), can be computed from either the time-derivative of $R(t)$ or the instantaneous shock position and velocity. We use the latter method because the former is sensitive to the spatial and temporal resolutions and is generally less precise. The time in Eq.\,(\ref{eq:alpha}) is defined as zero when the shock is at the center. Since we cannot simulate fluids at $r=0$, we approximate the simulation time as starting at zero upon initialization. This approximation improves as time proceeds. 

\section{Shocks in an $\MakeLowercase{n}=2.5$ Polytrope}
\label{sec:results1}

We take an $n=2.5$ polytrope (i.e., $\rho_1\propto r^{-2.5}$) as our fiducial model. This is both a typical polytropic index for which SS, CQR, and RW solutions exist, and is a good approximation for red and yellow supergiant hydrogen envelopes over a large range of radii (Paper I).

\subsection{Initial and Asymptotic Shock Evolution}
\label{sec:initialphase}
\begin{figure}[ht]
\centering
\includegraphics[width=\columnwidth]{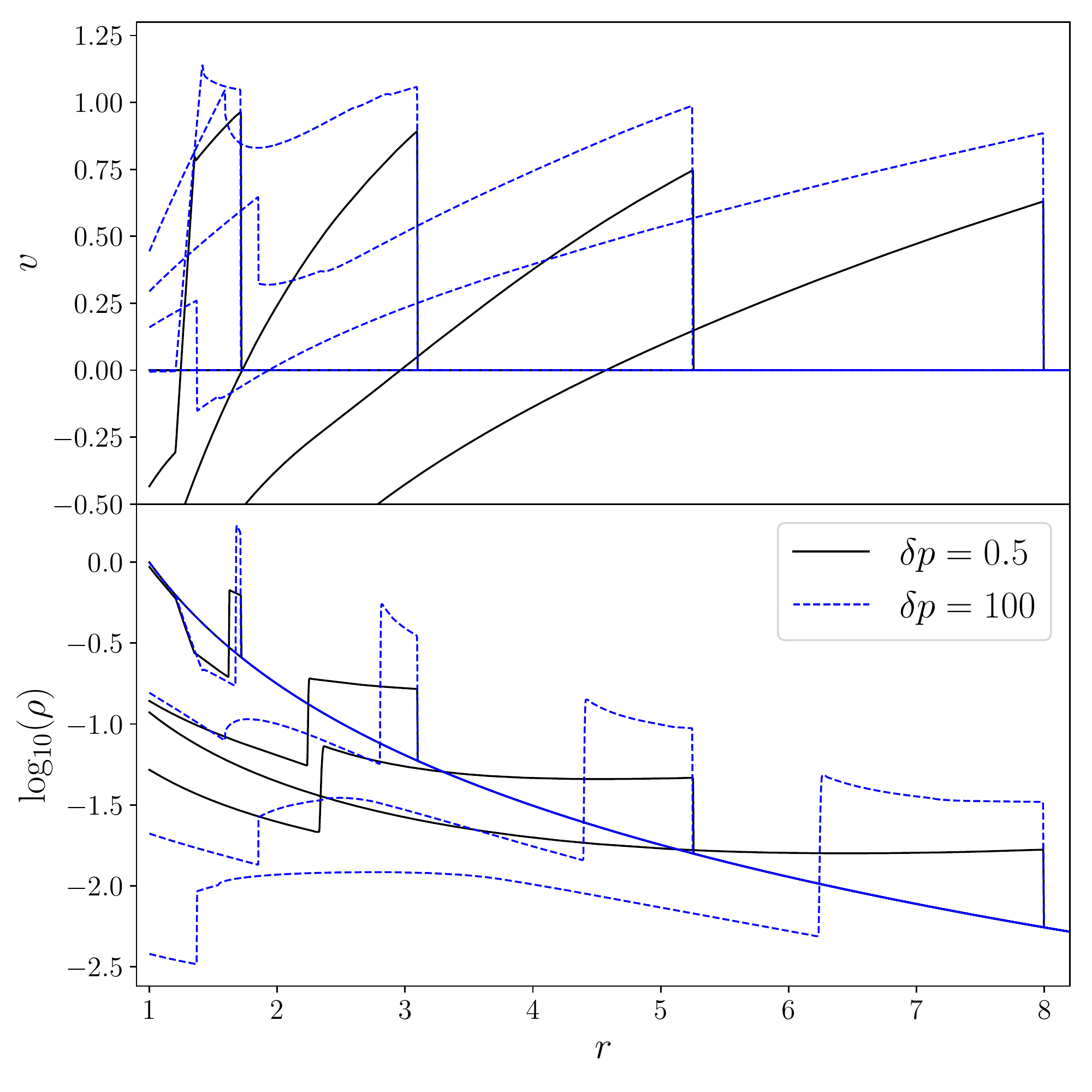}
\caption{Velocity and density structures of the early evolution of shocks with intermediate (black) and strong strengths (blue dashed) for $n=2.5$. Velocities behind the strong shock are divided by a factor of $10$. The shock is immediately in front of the velocity peak. The contact discontinuity is at the density `drop' behind the shock, which eventually leaves through the inner boundary for the shock with intermediate strength. Solutions for the different shock strengths are shown at the same shock radius, which corresponds to different times in each simulation.} 
\label{fig:timezoom}
\end{figure}

Fig.\,\ref{fig:timezoom} shows the early hydrodynamic evolution of a shock with intermediate ($\delta p=0.5$, $M_{\rm s}(t\!=\!10^2)\simeq2.2$) and high strength ($\delta p=100$, $M_{\rm s}(t\!=\!10^2)\simeq29$) in an $n=2.5$ polytrope. A shock forms at the over-pressurized edge, $r=1.5$, and expands outward in radius. A rarefaction wave forms behind the shock that propagates inward and leaves the inner boundary, which can be seen in the earliest density profile in Fig.\,\ref{fig:timezoom}. Separating the shocked and rarified materials is a contact discontinuity (CD), which propagates outward at a slower velocity than the shock. A second, weak rarefaction wave propagates outwards toward the shock from the inner boundary informing the fluid of the `black hole'. The CD is eventually engulfed by the outgoing rarefaction wave and descends toward the inner boundary. In simulations of stronger shocks, the shocked material is unaware of the black hole for a longer amount of time. Information of the black hole may never reach a very strong shock if either a hollow interior ($n\ge n_h$) or sonic point ($n>3$) forms first.

We find the post-shock distributions of physical quantities in our array of simulations do not qualitatively change after $t\ge10^2$. We label the logarithm of time between $t=10^2$ and $10^3$ as the `asymptotic' phase of the shock evolution. The distributions of shocked material in the last snapshot in Fig.\,\ref{fig:timezoom} are representative of the distributions seen in the asymptotic phase. For the strong shock in particular, the distribution above the CD remains qualitatively unchanged. 


\subsection{Suite of Shock Simulations}
\begin{figure}[ht]
\centering
\includegraphics[width=\columnwidth]{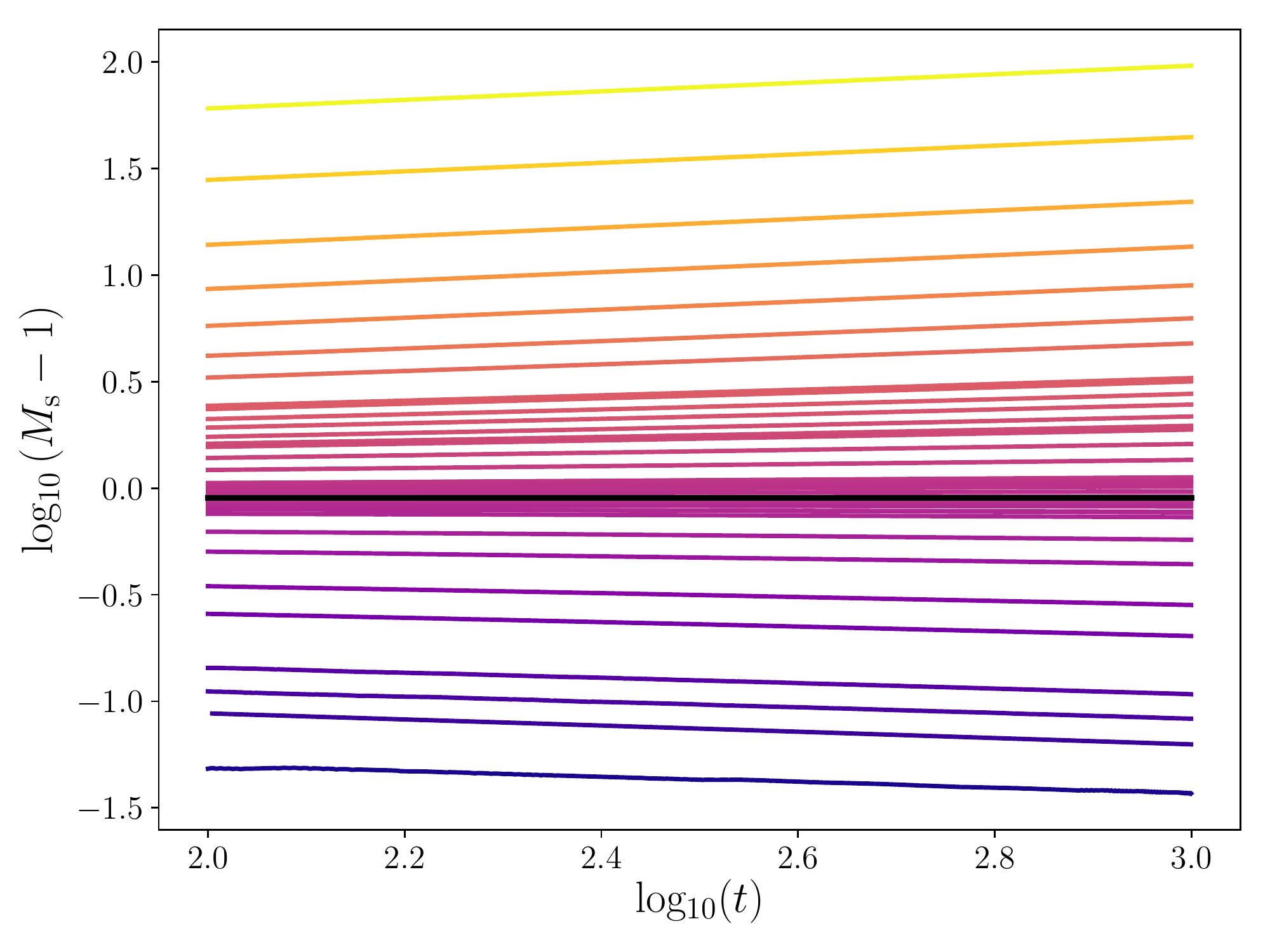}
\caption{Time evolution of the Mach number for shocks of varying initial strength in an $n=2.5$ polytrope. Each coloured line is one simulation. The horizontal black line is the predicted value, $M_{\rm CQR}\simeq1.90$, from Paper I. Shocks with Mach numbers above or below this predicted value monotonically strengthen or weaken with time, respectively. }
\label{fig:dmach_time}
\end{figure}

In Paper I, we predict the CQR solution for $n=2.5$ follows a constant Mach number of $M_{\rm CQR}\simeq1.90$ or log$_{10}(M_{\rm CQR}-1)\simeq-0.045$. Fig.\,\ref{fig:dmach_time} shows time evolution of the Mach number from simulations with each color labeling an initial condition of varying shock strength. The black horizontal line in Fig.\,\ref{fig:dmach_time} is the prediction from Paper I and independent of the simulation results. We find shocks with larger (smaller) Mach numbers than $M_{\rm s} \simeq1.90$ continue to strengthen (weaken) in time, consistent with the result of Paper II that the CQR solutions are weakly linearly unstable. This is also true for shocks with Mach numbers far from the CQR prediction where the linear analysis of the CQR solutions from Paper II  breaks down.

\begin{figure}[ht]
\centering
  \subfigure[$n=2.5$ ]{\includegraphics[width=\columnwidth]{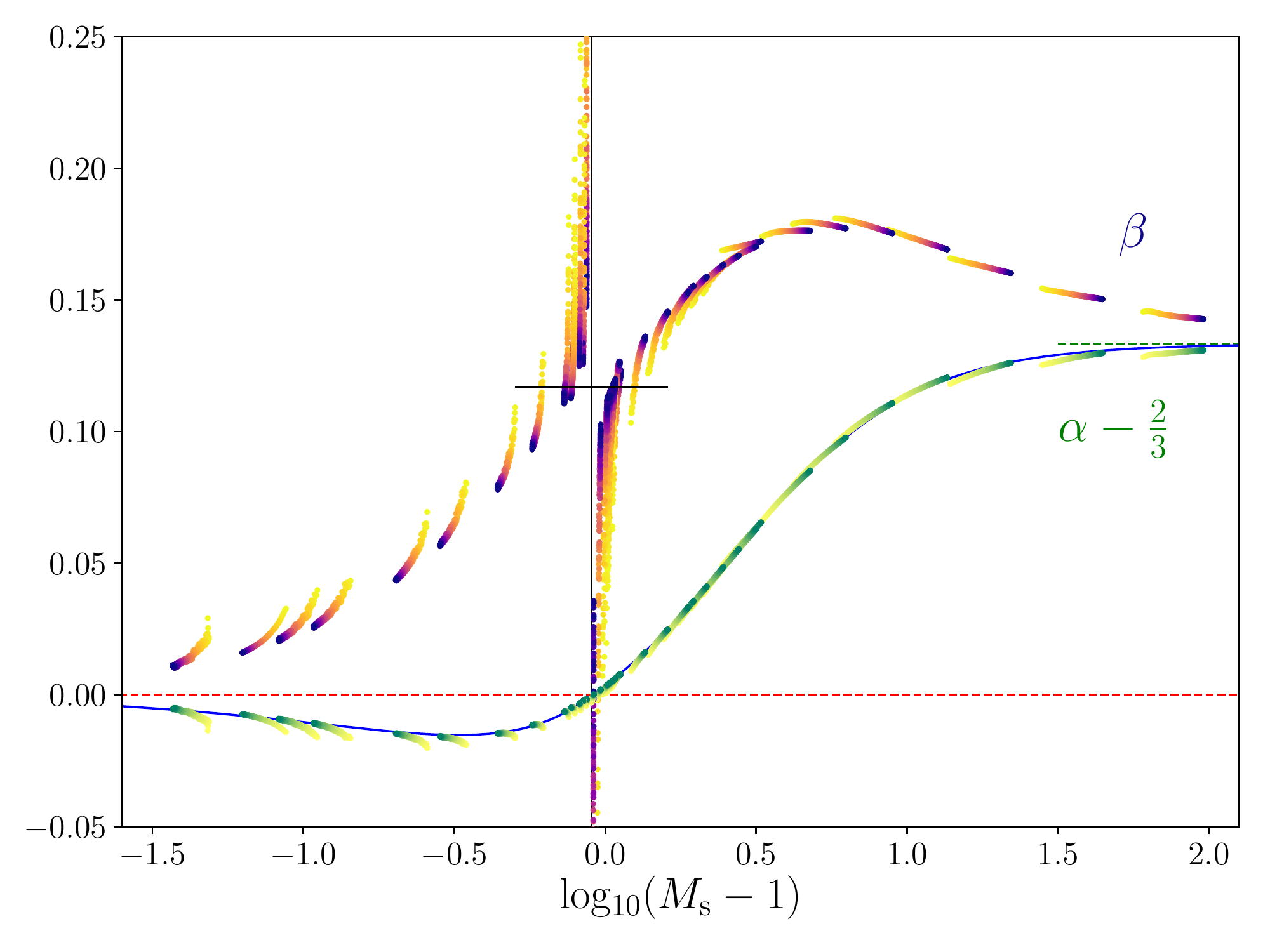}\label{fig:n2p5}}
  \subfigure[$n=2.9$ ]{\includegraphics[width=\columnwidth]{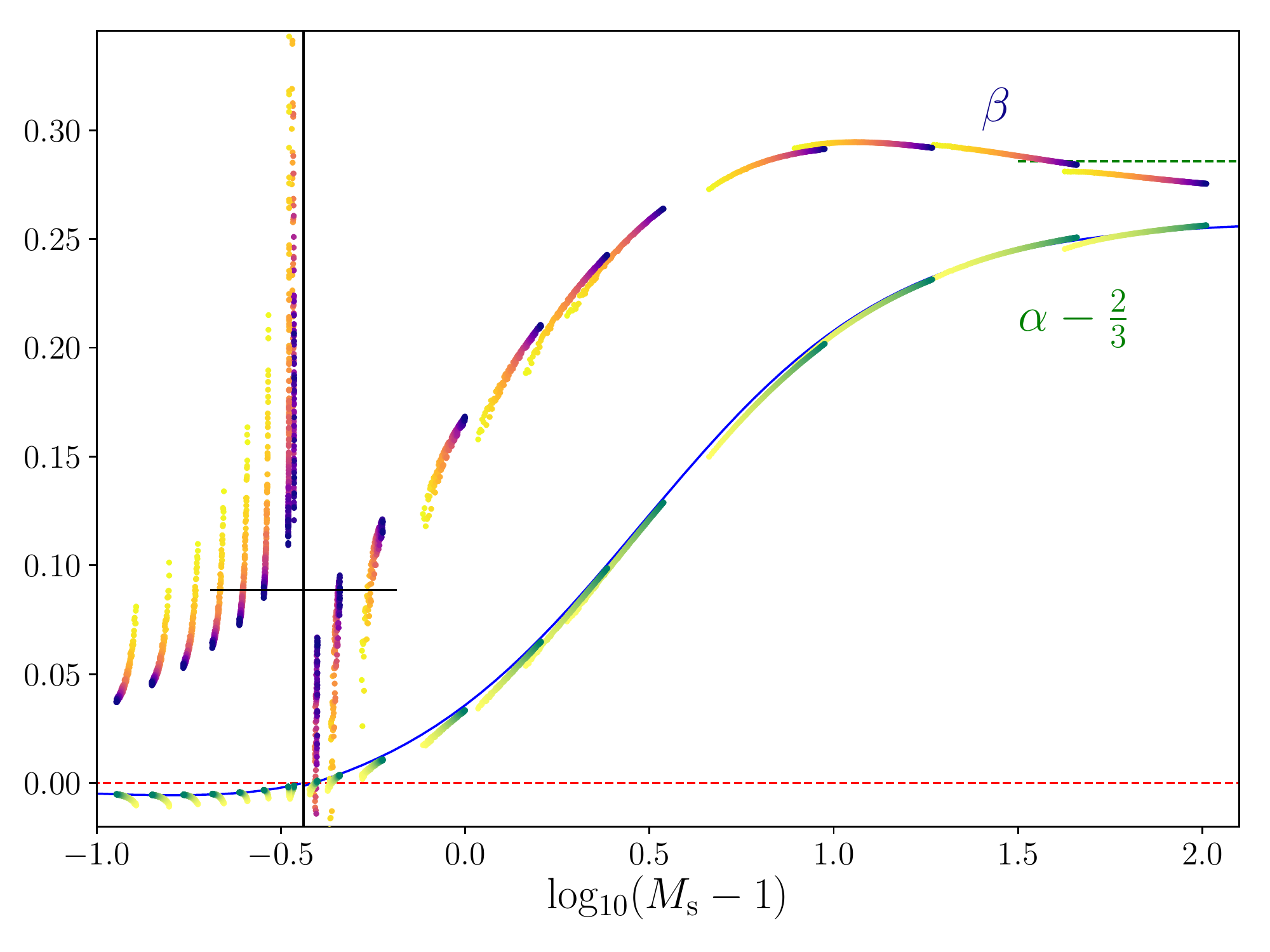}\label{fig:n2p9}}
\caption{Green/yellow points show the temporal power-laws, $\alpha=d{\rm ln}(R)/d{\rm ln}(t)$, of the shock trajectories with respect to a trajectory with constant Mach number: $\alpha-2/3$ (Eq.\,\ref{eq:alpha}). Purple/yellow points show the growth rate of the shock position with respect to the self-similar solution from Paper I (Eq.\,\ref{eq:sigma2}). Time proceeds from yellow to green/purple. Red dashed line is the predicted temporal power-law, $\alpha=2/3$, for the self-similar solutions for weak and infinitely weak shocks. Black vertical and horizontal lines are the predictions for $M_{\rm CQR}$ and $\beta_{\rm CQR}$ from Papers I \& II. Green dashed lines are the expected values for both $\alpha-2/3$ and $\beta$ in the Sedov-Taylor limit. The blue lines are empirical relations for $\alpha-2/3$ from Table\,\ref{table:2}. 
}
\label{fig:alphasigma_dmach}
\end{figure}  

\subsubsection{Shock Trajectories}
\label{sec:results_trajectory_n2.5}
Each cluster of yellow/green points in Fig.\,\ref{fig:n2p5} corresponds to a simulation spanning $t=10^2\!-\!10^3$, with yellow being the earliest time. Their values are the instantaneous temporal power-laws of the numerical simulations with respect to a trajectory with constant Mach number (i.e., $\alpha-2/3$). Shocks that strengthen (weaken) with time have a higher (lower) Mach number at the green end of the cluster of points. Fig.\,\ref{fig:n2p5} shows that shocks that strengthen with time have $\alpha>2/3$, while shocks that weaken with time have $\alpha<2/3$. Also shown as dashed lines are the predicted values of the RW, CQR, and SS temporal power-laws (\S\ref{sec:trajectory}). Infinitesimally weak shocks are effectively sound waves and, so, we expect and observe that $\alpha$ approaches $2/3$ for very small Mach numbers. Likewise, very strong shocks have temporal power-laws that agree with the predictions for Sedov-Taylor blast waves: $\alpha_{\rm ST}={4/5}$. 

The values for $\alpha$ and $M_{\rm s}$ from all of our simulations appear to follow a single contour, $\alpha(M_{\rm s})$, that limits to the Sedov-Taylor and RW values. This contour smoothly passes through $\alpha=2/3$ at exactly the Mach number predicted for a CQR solution. The solid blue lines in Fig.\,\ref{fig:alphasigma_dmach} are empirical relations from Table\,\ref{table:2} in Appendix C.

All of the calculations of $\alpha$ in Fig.\,\ref{fig:alphasigma_dmach} span a factor of 10 in time, $t=10^2-10^3$. Clusters of points near the CQR Mach number span very narrow ranges of $\alpha$ and $M_s$ in this time in comparison to shocks with much larger or smaller Mach numbers. Therefore, shocks near the CQR solution evolve very slowly. In Paper II, we find that the linearly unstable growing mode affects the shock trajectory of a perturbed CQR solution in the following manner:
\beq
R(t)\simeq R_{\rm CQR}(t)\left(1+\zeta' (t-t_0)^{2\sigma/3}\right), \label{eq:rshock_perturbedcqr}
\eeq
where $R_{\rm CQR}(t)$ is the unperturbed CQR solution (Eq.\,\ref{eq:rshock_cqr}) and $\zeta'$ encodes the perturbation amplitude. The growth rate, $\sigma(n,\gamma)$, takes on a characteristic value for a given $n$ and $\gamma$. We can compute the growth rate from our simulations using the following expression:
\bea
\beta&\equiv&\frac{d{\rm log}}{d{\rm log}(t)}\left(\frac{R(t)}{R_{\rm CQR}(t)}-1 \right) \label{eq:sigma} \\
&=& \left(\alpha- \frac{2}{3}\right) \left(1-\frac{R_{\rm CQR}(t)}{R(t)} \right)\inv,
\label{eq:sigma2}
\eea
and compare this to the predicted growth rate from Paper II. The second equation above is derived with the knowledge that $\alpha_{\rm CQR}=2/3$. From here on, we refer to the predicted values of $\sigma$ as $\beta_{\rm CQR}=2\sigma/3$. For $n=2.5$, these values are $\sigma\simeq0.175$ and $\beta_{\rm CQR}\simeq0.117$. 

The yellow/purple cluster of points in Fig.\,\ref{fig:n2p5} shows Eq.\,(\ref{eq:sigma2}) as a function of Mach number. The shock location for the unperturbed CQR solution, Eq.\,(\ref{eq:rshock_cqr}), is defined by the numerical shock position at time $t_0=1$. The second factor on the right side of Eq.\,(\ref{eq:sigma2}) approaches either zero or one if the shock is very weak or strong. Thus, we expect $\beta$ is zero in the RW limit and $\alpha_{\rm ST}-2/3\simeq 0.133$ in the Sedov-Taylor limit. Indeed, our numerical results agree with these analytic expectations. We also see that shocks with Mach numbers near the CQR value have trajectories that grow with the expected value of $\beta_{\rm CQR}\simeq0.117$. This result is consistent with our detailed analysis in Paper II of high-resolution simulations of shocks with Mach numbers near the CQR value. The similarity between the values of $\beta_{\rm CQR}\simeq0.117$ and $\alpha_{\rm ST}-2/3\simeq0.133$ is a coincidence, as demonstrated explicitly in Fig.\,\ref{fig:n2p9}, which shows the shock trajectories in an $n=2.9$ polytrope.

From Papers I and II, we expect the following properties of the CQR solution for $n=2.9$: $M_{\rm CQR}\simeq1.36$ or log$_{10}(M_{\rm CQR}-1)\simeq-0.44$, $\beta_{\rm CQR}\simeq0.09$, and $\alpha_{\rm ST}-2/3\simeq0.29$. The numerical results in Fig.\,\ref{fig:n2p9} are reasonably consistent with the analytics.  Fig.\,\ref{fig:n2p9} does show a small difference between the hydrodynamic simulations and theoretical predictions toward the Sedov-Taylor limit. This discrepancy is related to the formation of a hollow interior with an infinite density at the CD, which is numerically difficult to model, as is discussed further in Appendix B.

\subsubsection{Post-Shock Solutions}
\label{sec:postshocksolution}
\begin{figure}[]
  \centering
    \subfigure[Velocity]{\includegraphics[width=0.95\columnwidth]{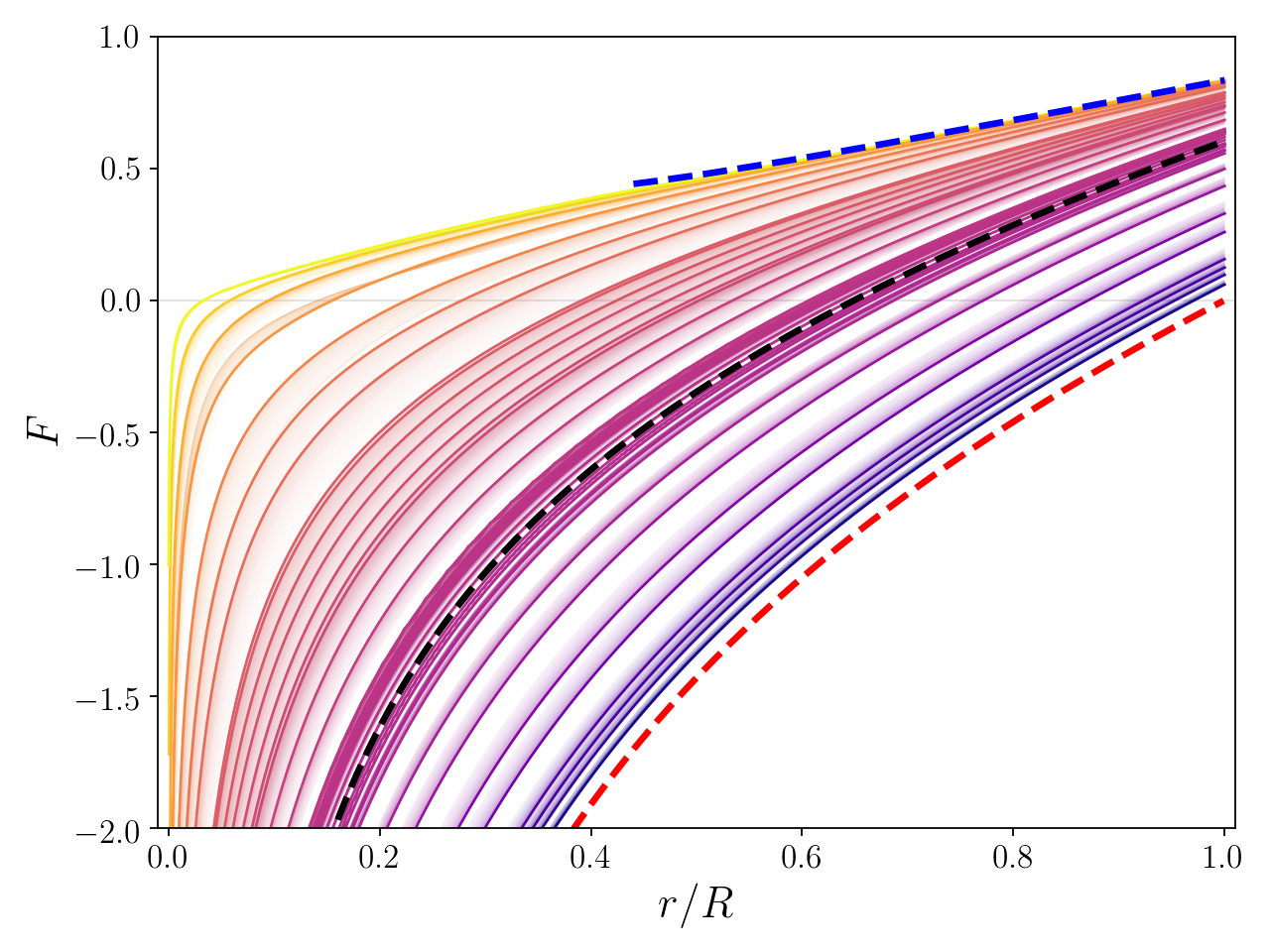}}
    \subfigure[Density]{\includegraphics[width=0.95\columnwidth]{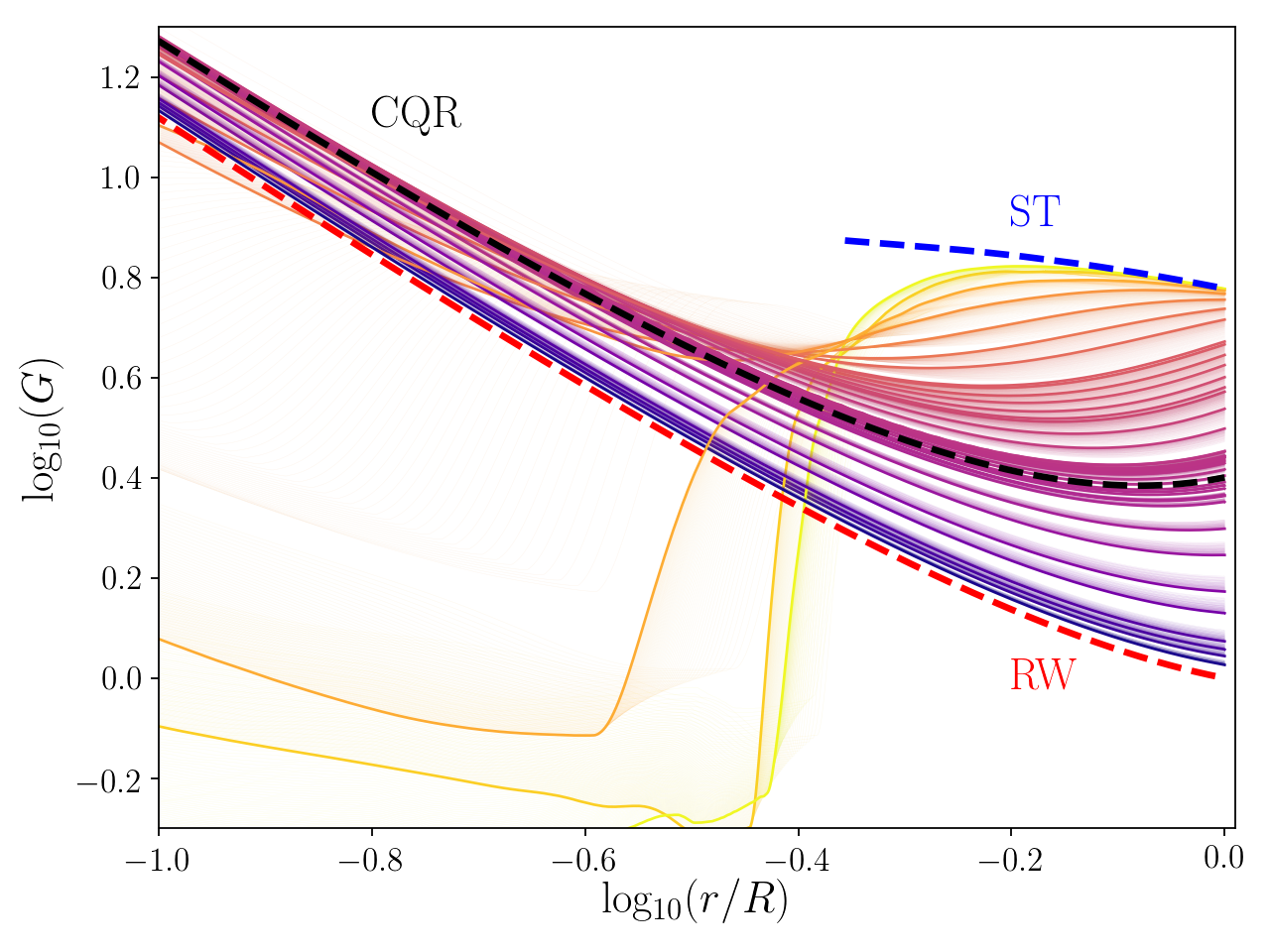}}
    \subfigure[Pressure]{\includegraphics[width=0.95\columnwidth]{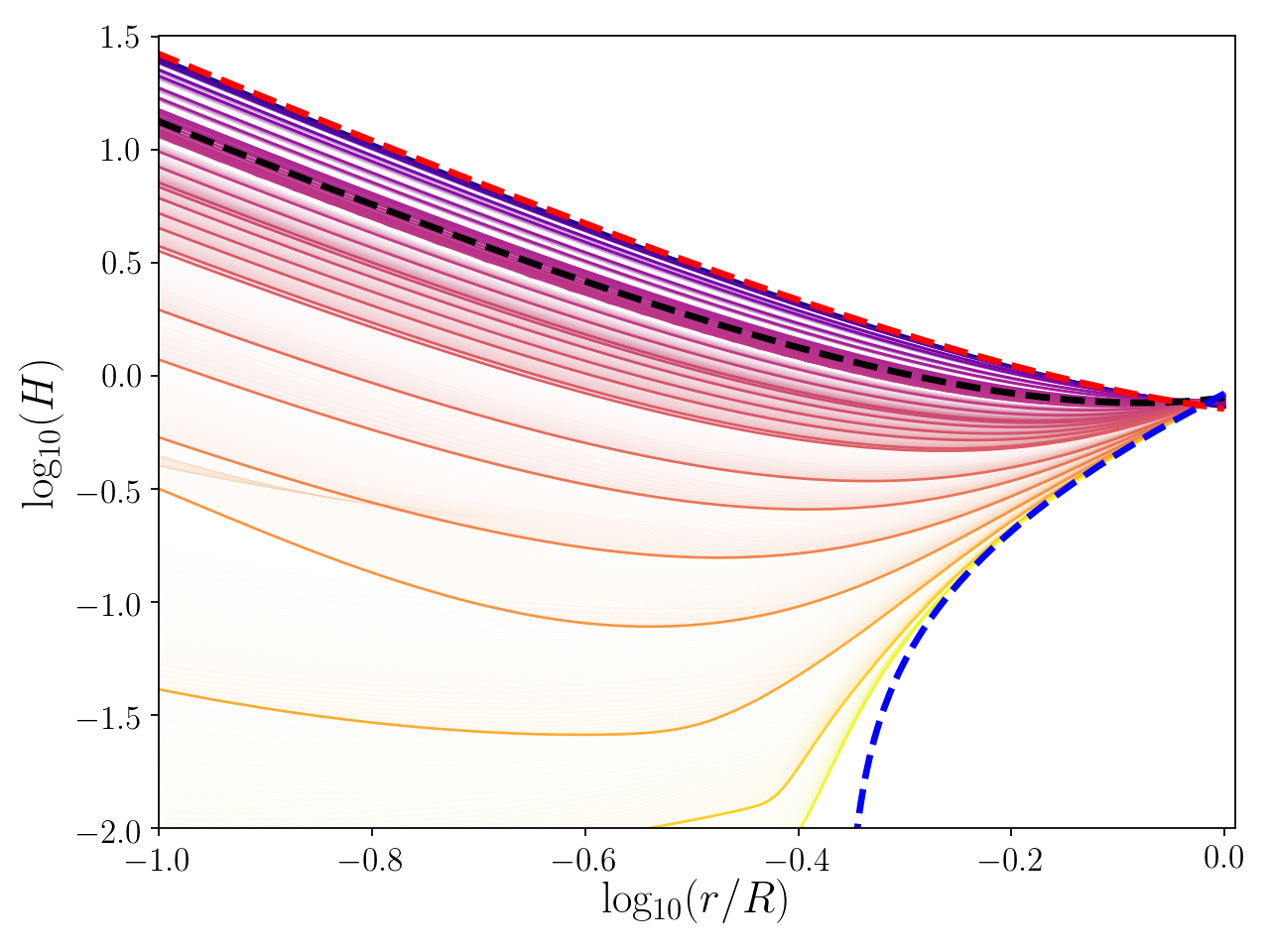}} 
  \caption{Post-shock solutions for $n\!=\!2.5$ between $10^2\le\!t\!\le10^3$ (i.e., velocity [$F$], density [$G$], and pressure [$H$]; Eq.\,\ref{eq:dimensionless}). Thin lines represent an instance in time sampled at every $\Delta t\!=\!5$. The thickest line is the latest time, $t\!=\!10^3$. Solid lines have corresponding colours to Fig.\,\ref{fig:dmach_time}. Red, black, and blue dashed lines are the self-similar solutions for infinitely weak, weak, and infinitely strong shocks, respectively, for $n\!=\!2.5$. Note, the Sedov-Taylor solution is hollow (i.e., a vacuum) below $r/R\!\lesssim\!0.44$.}
  \label{fig:fgh}
\end{figure}

Fig.\,\ref{fig:fgh} shows the post-shock solutions for self-similar RW, CQR, and Sedov-Taylor solutions. This figure adopts the definitions from Paper II for the velocity, density, and pressure variables: 
\beq
f(\xi) = \frac{v}{V(t)}, \ \ g(\xi) = \frac{\rho}{\rho_1(R)}, \ \ h(\xi) = \frac{p}{\rho_1(R)V(t)^2}
\label{eq:dimensionless}
\eeq
where $\xi=r/R(t)$ is the dimensionless radial coordinate defined between the shock ($\xi=1$) and black hole ($\xi=0$). Each color in Fig.\,\ref{fig:dmach_time} and \ref{fig:fgh} represents the same simulation. The collection of thin lines in Fig.\,\ref{fig:fgh} are the post-shock solutions from our numerical simulations between $t=10^2\!-\!10^3$, which uniformly sample time every $\Delta t=5$. The thickest line is the solution at $t=10^3$.

Populating the gaps between the three self-similar solutions are the post-shock solutions from our suite of simulations. We see shocks with Mach numbers close to the CQR value have post-shock solutions that resemble the CQR solution. This resemblance is expected and discussed in detail in Paper II. The post-shock solutions behind the weakest shock also resemble the RW solution. For small $\xi$, the post-shock solutions from all of our non-strong shock simulations (i.e., $M_{\rm s} \lesssim3$) accrete in a Bondi-fashion since the dynamics there are predominantly set by the black hole. The post-shock solution behind the strongest shock resembles the Sedov-Taylor solution above the CD. As time proceeds, we see the pressure and density below the CD decreases to, presumably, form a hollow interior. 


These solutions show that shocks weaker than CQR have post-shock solutions that evolve away from the CQR solution and toward the RW solution. Likewise, the post-shock solutions behind shocks stronger than CQR evolve toward the Sedov-Taylor solution. Moreover, a comparison to Fig.\,\ref{fig:dmach_time} shows the order of these solutions are sequential with Mach number, which itself grows monotonically with time. The trend is seen for each individual simulation and also across the suite of simulations, and suggests our suite of simulations is not solely a collection of individual explosions. Rather, each simulation (at late times) samples a decade in time of a single explosion's long term evolution. 

One puzzle raised by our results is related to the formation of a hollow interior in the Sedov-Taylor solution. The strongest shock in Fig.\,\ref{fig:fgh} is the late evolution of the strong shock in Fig.\,\ref{fig:timezoom}. The CD in the density distribution from Fig.\,\ref{fig:timezoom} is preserved at late times and rests near the predicted self-similar radial coordinate $\xi_{\rm CD}\simeq0.44$ or log$_{10}(\xi)\simeq-0.36$ for $n=2.5$. If a weaker shock that has lost its original CD strengthens toward the Sedov-Taylor limit with time, how does the solution develop another CD? Our simulations do not span a long enough time to answer this question.

\section{Shocks in Polytropes}
\label{sec:results2}
\begin{figure*}[ht]
\centering
\setbox1=\hbox{\includegraphics[width=\textwidth]{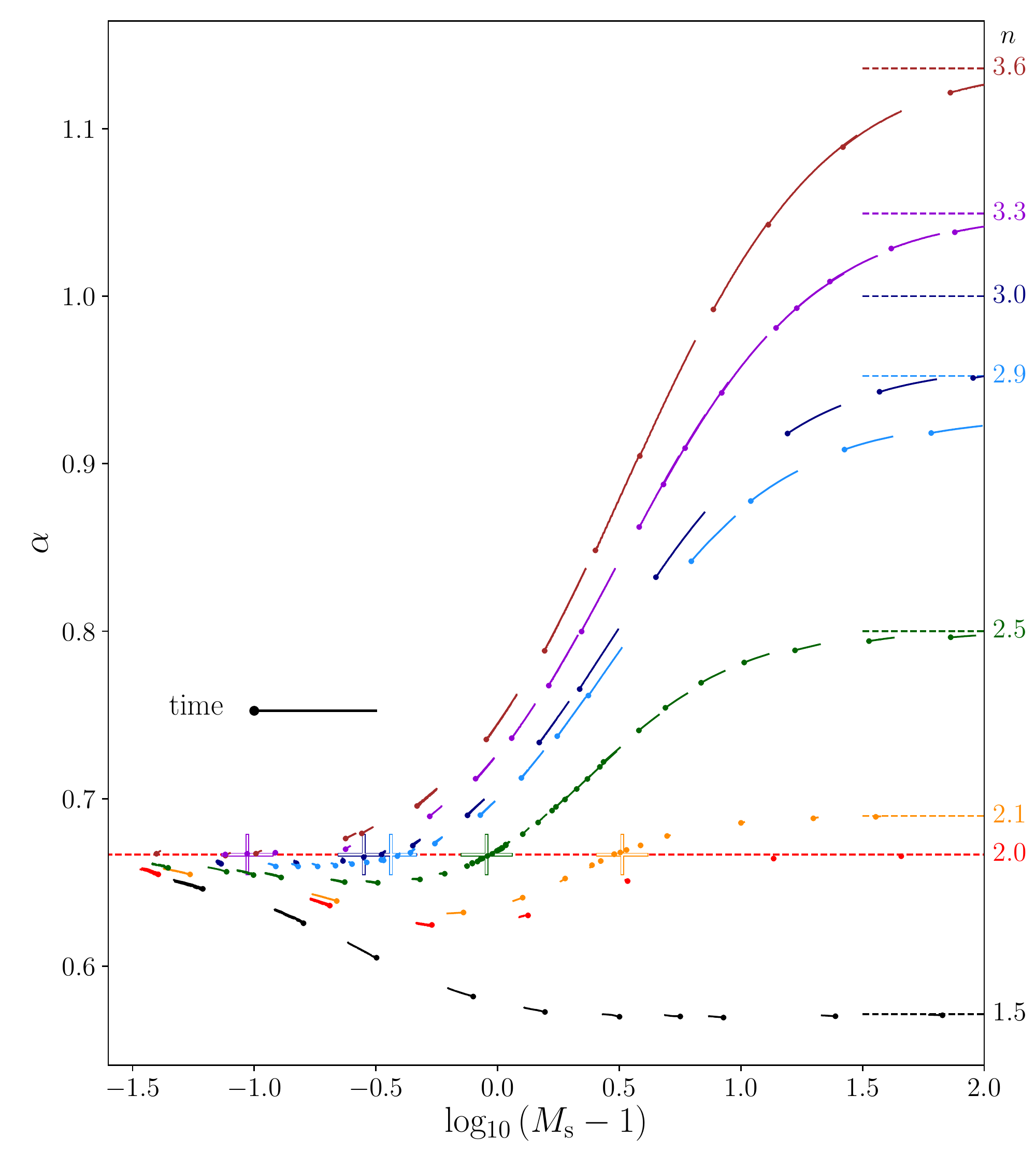}}
\includegraphics[width=\textwidth]{mega_alpha_mach_cut250sec.pdf}\llap{\makebox[\wd1][l]{\hspace{16.5ex}\raisebox{0.495\textheight}{\includegraphics[width=0.425\textwidth]{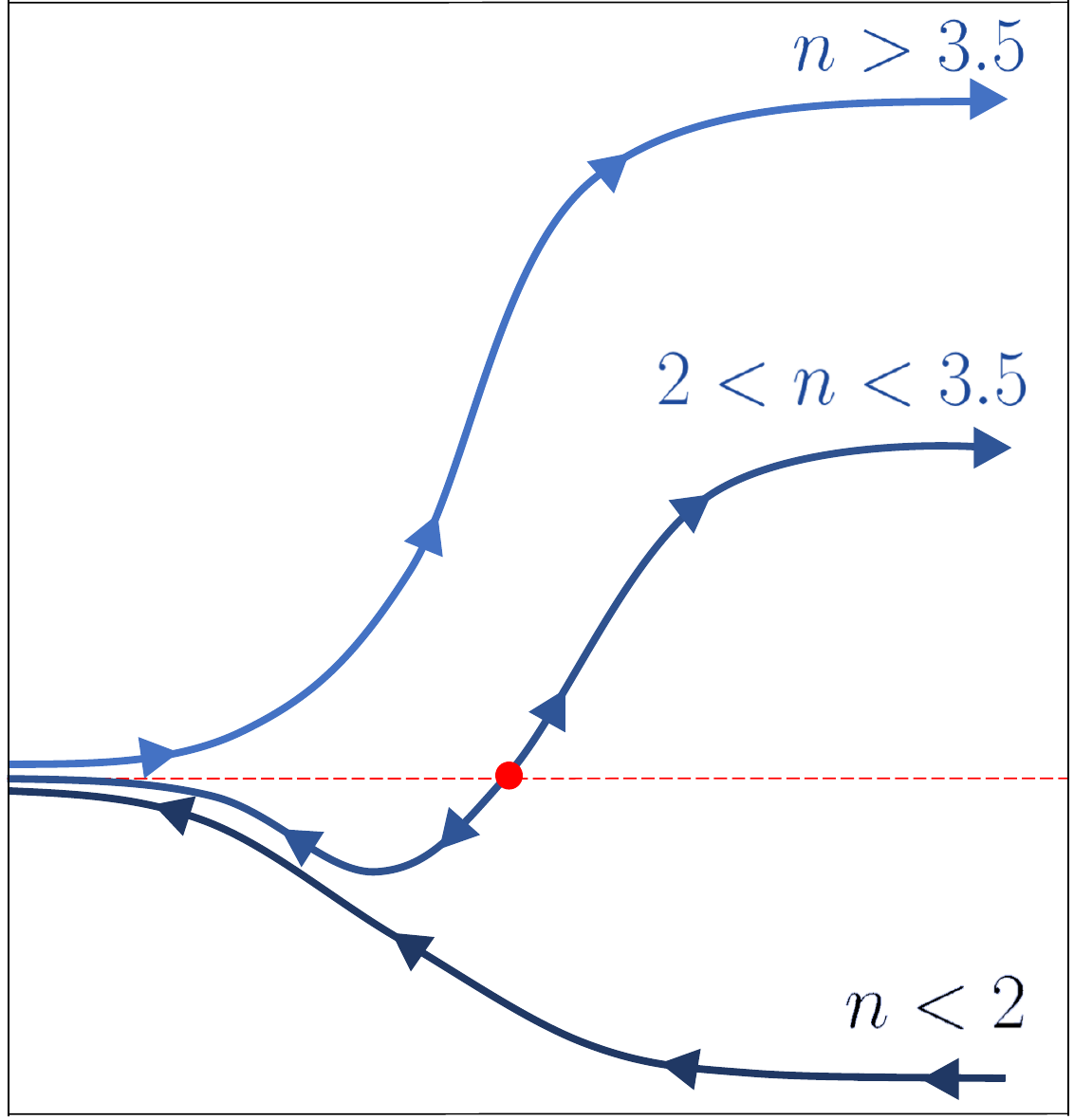}}}}
\caption{Each solid line is the temporal power-law of a shock trajectory, $\alpha=d{\rm ln}(R)/d{\rm ln}(t)$, as a function of Mach number for $t\ge250$. The dot represents $t=250$ so that the direction of each line relative to the point shows if the Mach number increases or decreases with time. Colours correspond to a polytropic index labeled in the right margin. Dashed lines are the self-similar values. Coloured crosses are the solutions for weak self-similar shocks from Paper I. The inset is a schematic of how shocks evolve in time depending on the initial strength and ambient density profile. For $n\le2$ ($n\ge3.5$), shocks in our simulations weaken (strengthen) with time and follow a curve toward the values for infinitely weak (strong) self-similar shocks. For $2<n<3.5$, the weak self-similar solution from Paper I is an unstable equilibrium point (red dot). Shock trajectories migrate away from this solution towards either the infinitely weak or strong limits.} 
\label{fig:mega_alpha}
\end{figure*}  


\subsection{Phase Portrait of Shock Trajectories}
\label{sec:megaalpha}
Our results in \S\ref{sec:results1} suggest that the numerical simulations of explosions of different strength in an $n=2.5$ polytrope are all closely related to each other. Each simulation (at late times) effectively samples the long term evolution of a single explosion. The state of this explosion is described by two `phase variables': the shock Mach number, $M_s(t)$, and temporal power-law, $\alpha(t)$. In \S\ref{sec:results_trajectory_n2.5}, we found the value $\alpha(t)$ measures whether $M_s(t)$ increases or decreases with time. This suggests $M_s(t)$ and $\alpha(t)$ are the `phase position' and `phase velocity' of an explosion .
Another example of where this nomenclature is used is the simple pendulum problem where the angular position, $\theta(t)$, and velocity, $\dot{\theta}(t)$, are the phase variables for the pendulum's state. Fig.\,\ref{fig:alphasigma_dmach} is an example of a `phase portrait' or a graph of the phase variables for an explosion.

Fig.\,\ref{fig:mega_alpha} shows a phase portrait of the shock trajectories from all of our simulations including those for $n=2.5$ for $t\ge250$. The lines indicate the direction in which $\alpha$ is changing with time. Broadly, we see that all shocks with $\alpha<2/3$ weaken with time (i.e., the Mach number decreases). Likewise, all shocks with $\alpha>2/3$ strengthen in time (i.e., the Mach number increases). These results support the suggestion from \S\ref{sec:trajectory} that $\alpha-2/3$ is a measure of the growth of a shock's strength.

More specifically, we see that all shocks, including those that are initially strong, in $n\le2$ polytropes decay toward the RW solution, while for $n\ge3.5$ all shocks -- including those that are initially weak -- strengthen toward the SS solution from \cite{ws93}. For $2<n<3.5$, all shocks with Mach numbers larger (smaller) than the CQR value strengthen (weaken) with time. For $n=2.9$ and 3.0, there is a small discrepancy in $\alpha$ for high Mach numbers related to the formation of a CD with infinite density; we discuss this further in Appendix B.

\subsubsection{Phase Portrait of Stable Equilibrium Solutions}

The system of hydrodynamic equations with boundary conditions defined by a shock and sonic point (due to the black hole) describes a dynamical system. Our investigation studies the solutions to this dynamical system that start from an array of initial conditions. In \S\ref{sec:postshocksolution}, we find that the post-shock solutions evolve in a monotonic fashion toward a self-similar solution. Fig.\,\ref{fig:dmach_time} shows that the post-shock evolution varies monotonically with the shock Mach number, which also grows monotonically at late times. In \S\ref{sec:results_trajectory_n2.5} and \S\ref{sec:megaalpha}, we show a representation of a shock's space-time trajectory in terms of the temporal power-law, $\alpha$, and Mach number, $M_{\rm s}$. At late times, the shock trajectories appear to lose memory of their initial conditions and collectively rest on a single curve in $\alpha$ and $M_s$. This suggests that the system has a stable equilibrium solution, $\alpha_n(M_{\rm s})$, for each $n$ to which shock trajectories are attracted and follow at late times. In Fig.\,\ref{fig:mega_alpha}, we embed a schematic phase portrait of the three types of stable equilibrium solutions suggested by our numerical and analytical results.

At high and low Mach numbers, the function, $\alpha_n(M_{\rm s})$, limits to the RW and SS solutions. For $2<n<3.5$, there is a crossing where $\alpha_n(M_{\rm s})=2/3$ at exactly the predicted Mach numbers from Paper I. If there were another self-similar solution besides RW, CQR, and SS then there must be five (or any odd number larger than three) total number of self-similar solutions. Otherwise, the weak and strong limits of $\alpha_n(M_{\rm s})$ will not reach the RW or SS values. The lack of two additional crossings in our numerical experiments in Fig.\,\ref{fig:mega_alpha} suggest \textit{there are no other self-similar solutions besides the RW, CQR, and SS solutions for this physical problem}. This statement is limited to the domain of Mach numbers simulated here. 

For $n<2$, the stable equilibrium solution is restricted to values of $\alpha_n<2/3$. The Sedov-Taylor solution is an unstable asymptotic limit from which all shock trajectories, no matter how strong initially, migrate away as $t\rightarrow\infty$. Because all of the trajectories approach the RW solution, we refer to this limit as asymptotically stable. Inverted behavior is seen for $n>3.5$: the stable equilibrium solution is restricted to values of $\alpha_n>2/3$ and the stability of the asymptotic limits reverse. All shocks, including initially weak shocks, have trajectories that migrate away from an unstable asymptotic limit, the RW solution, toward a stable asymptotic limit, the solution by \cite{ws93}. 

For intermediate $2<n<3.5$, an equilibrium point divides the stable equilibrium solution into two parts. Low Mach numbers have properties of the stable equilibrium solution for $n<2$, while high Mach numbers have the properties for $n>3.5$. The equilibrium point at intermediate Mach number is exactly the CQR solution found in Paper I. Our linear perturbation analysis in Paper II is, therefore, a study of trajectories near this equilibrium point. In our simulations, the equilibrium point is unstable since it repels all shock trajectories toward the stable asymptotic limits (RW or SS) at late times (e.g., Fig.\,\ref{fig:mega_alpha}). 

A relationship of the form $\alpha_n(M_{\rm s})$ suggests Eq.\,(\ref{eq:alpha}) is an implicit ordinary differential equation for the shock space-time trajectory, $\mathcal{R}(t)$:
\beq
\frac{t\dot{\mathcal{R}}}{\mathcal{R}} = \alpha_n\left( \frac{\dot{\mathcal{R}}}{c_1(\mathcal{R})} \right),
\label{eq:ode}
\eeq
where $\dot{\mathcal{R}}=\mathcal{V}$ is the shock velocity. Given the initial shock Mach number and position, we can integrate Eq.\,(\ref{eq:ode}) and solve for the shock trajectory and velocity, $\mathcal{V}_n(\mathcal{R})$, for each $n$. The strong shock solutions from \cite{1999ApJ...510..379M} take the form $\mathcal{V}(R,\rho)$ which has the same functional relation. This suggests there exists a weak extension to the strong shock solution by \cite{1999ApJ...510..379M} that includes a black hole. In Appendix C, we provide empirical relations of our numerical solutions for $\alpha_n(M_{\rm s})$ in Fig.\,\ref{fig:mega_alpha}.

\subsection{Accretion of Weakly Shocked Gas Onto a Black Hole}
\label{sec:mdot}
The rate of mass swept up by a self-similar shock with finite Mach number is proportional to the accretion onto a black hole, $\dot{M}$. Therefore, we expect the accretion rate to scale as
\beq
\dot{M}=4\pi R^2 \rho_1(R) V \propto t^{1 - 2n/3},
\eeq
where the right proportionality is derived using Eq.\,(\ref{eq:rho_ambient}), (\ref{eq:alpha}), and (\ref{eq:r_selfsimilar}). This is also the accretion rate for a RW solution for all $n$ since it is a `shock' with a Mach number of one. The accretion rate for a RW solution grows with time for $n<3/2$ and declines otherwise. Since the CQR solution exists only for $2<n<3.5$, the accretion rate always declines with time. 

\begin{figure}[]
\centering
\includegraphics[width=\columnwidth]{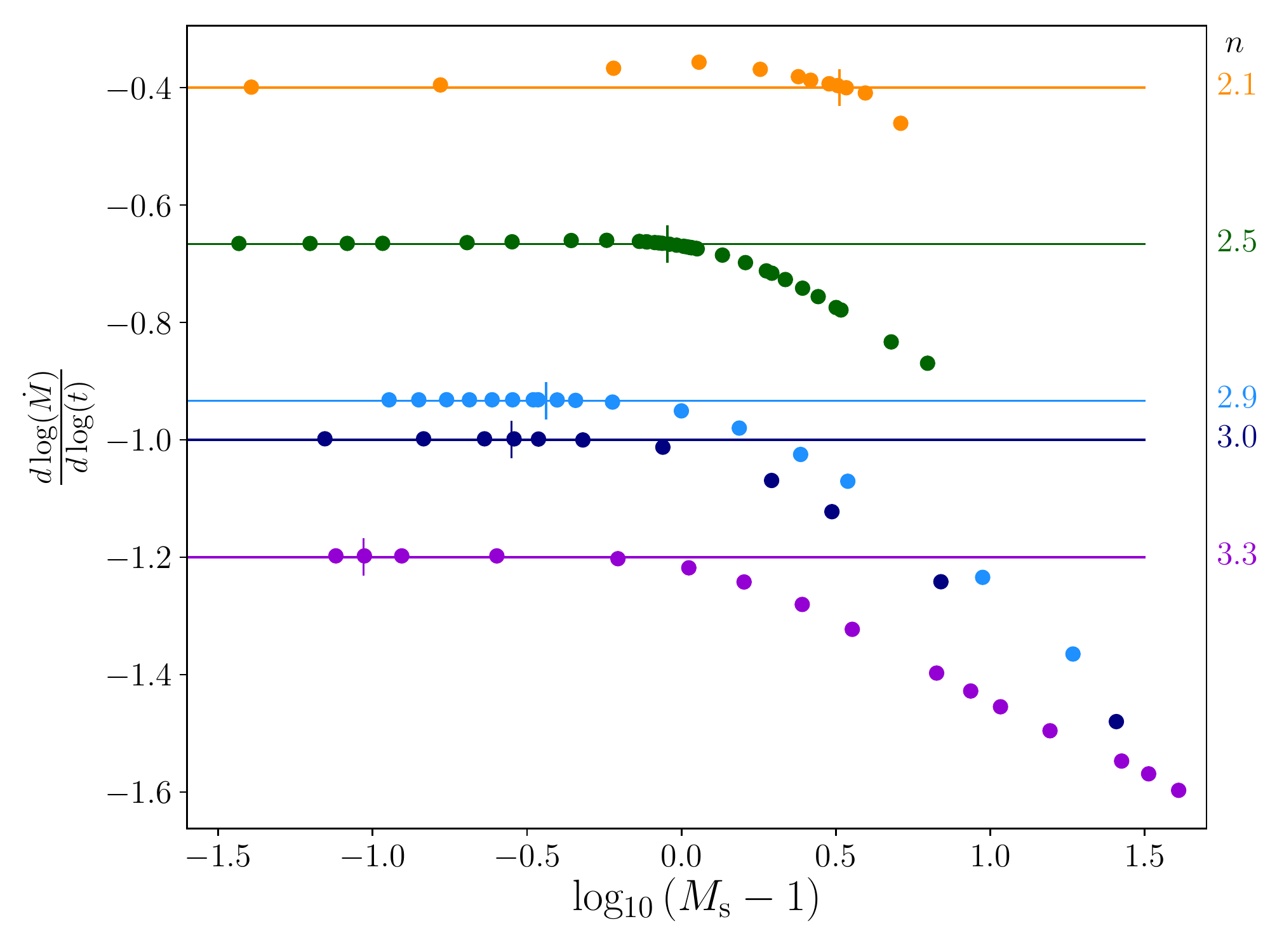}
\caption{Each point is the temporal power-law of the accretion rate, measured at the inner boundary, linearly-fitted between $500\le t\le10^3$. Each colour corresponds to a polytropic index labeled in the right margin. Horizontal lines are the predicted values, $1 - 2n/3$, and the vertical lines mark the Mach numbers for solutions for weak self-similar shocks from Paper I. For strong shocks, accretion is suppressed with respect to the values for free-fall or weak self-similar shocks because much of the mass is unbound. Simulations with a contact discontinuity in the domain are not included.}
\label{fig:mdot_dmach}
\end{figure}

Fig.\,\ref{fig:mdot_dmach} shows the fitted temporal power-law exponent of the accretion rate at the inner boundary of the simulations for $2<n<3.5$. The fit is made for $t\ge500$. The figure presents only the simulations where the contact discontinuity has been lost to the inner boundary. We exclude the accretion rates of shocks that retain their CD because the material below the CD is symptomatic of the initial conditions, which is not representative of realistic core-collapse. 

We see the power-law exponent is nearly constant for shocks with Mach numbers below or comparable to the CQR solutions. A small maximum exists slightly below the CQR value which corresponds to the minima in $\alpha$ in Fig.\,\ref{fig:mega_alpha}. The accretion rate declines steeply with time for shocks stronger than the CQR value since the shocked material is less bound and follows longer paths, in a Lagrangian sense, before falling into the black hole. The temporal exponent appears to decrease linearly with log$_{10}(M_{\rm s}-1)$ for shocks stronger than the CQR solution. Assuming the temporal exponent is a function of the shock Mach number,
\beq
\frac{d{\rm log}(\dot{M})}{d{\rm log}(t)} = f_n(M_{\rm s}-1),
\eeq
for each polytropic index $n$, the effective accretion rate could be estimated given a solution for Eq.\,(\ref{eq:ode}).

\section{Summary and Conclusions}
\label{sec:conclusion}

Simulations of core-collapse supernovae predict a large range of explosion energies. Weak explosions with energies comparable to, or less than, the stellar binding energy generate little ejecta. The majority of the shocked gas, instead, falls back onto the natal neutron star or black hole on timescales that depend on the finite shock strength, finite sound speed of the progenitor, and local escape speed. For these reasons, the Sedov-Taylor solution is not applicable for weak explosions. 
For certain stellar configurations, however, there are other self-similar solutions that describe weak explosions. 

In stellar supergiants, the gravitational field in the envelope is similar to a point mass since the stellar mass is so centrally concentrated. As a result, radial profiles of the envelope density and pressure follow simple power-laws (i.e., polytropic profiles). Under this configuration, we derived two sets of self-similar solutions in Papers I and II. These solutions describe weak and infinitely weak explosions. Unlike the Sedov-Taylor solution, these do not have a constant energy in the post-shock gas because they account for both the energy of the envelope that passes through the shock and accretion induced by core-collapse. We label the three self-similar solutions as SS, CQR, and RW in reference to the strong shock (e.g., Sedov-Taylor), weak shock, and rarefaction wave solutions, respectively. The RW and SS solutions exist for density profiles, $\rho_1\propto r^{-n}$, with any polytropic index, $n$, and correspond to either an infinitely weak or strong shock. The CQR solution has a finite characteristic Mach number and only exists between $2<n<3.5$. All of the self-similar solutions depend on the adiabatic indices, $\gamma_i$, of the pre- and post-shock gas (i.e., $p\propto \rho^{\gamma_i}$). To simplify our study here, we assume the adiabatic indices are identical and related to the polytropic index in the form $\gamma=1+n\inv$. This is a good approximation for low energy explosions in supergiants (\citealt{2018MNRAS.477.1225C}, Paper I). In high energy explosions, however, $\gamma_2=4/3$ in the post-shock gas and is independent of the pre-shock conditions.

One application of all of these solutions is a failed supernova, in which the bounce shock from the neutron star is not revived. Instead, the liberation and (nearly instantaneous) escape of $\sim few\times0.1\,\msun$ of mass-energy in neutrinos from the protoneutron star results in an over-pressurized stellar envelope. Consequently, the envelope expands and triggers a global acoustic pulse. The acoustic pulse either steepens into a weak shock or damps to form a rarefaction wave (i.e., an infinitely weak shock). We expect the self-similar solutions are good representations of shocks with Mach numbers close or equal to a self-similar value. In general, however, a failed supernova shock can have any Mach number not equal to a self-similar value. In this paper, the third in a series, our goal has been to understand the role of self-similar solutions in the long term evolution of shocks that are not self-similar. We employed hydrodynamic simulations to study explosions in polytropic environments with a central point mass. We presented a suite of simulations that samples a wide range of explosion energies and polytropic indices.

An explosion generates a shock that propagates along a trajectory, $R(t)$, in space-time. Expressing the trajectories in terms of the shock Mach number, $M_s$, and instantaneous temporal power-law, $\alpha=t\dot{R}/R$, provides a compact summary of the shock properties. Phase portraits of these trajectories are shown in Fig.\,\ref{fig:alphasigma_dmach} and \ref{fig:mega_alpha}. All shocks are seen to either weaken or strengthen with time for $n<2$ or $n>3.5$, respectively. Very strong or weak shocks have temporal power-laws that correspond to the SS and RW solutions, respectively.\footnote{For $n=2.9$ and 3, there is a small discrepancy between our numerics and known analytics for strong shocks due to the slow formation of a hollow interior with an infinitely dense contact discontinuity. See Appendix B.} Between $2<n<3.5$, we find all shocks with Mach numbers stronger (weaker) than the CQR value continue to strengthen (weaken) with time. Shocks with Mach numbers and temporal power-laws near the CQR solution grow away from the CQR solution at an extremely slowly rate. Therefore, these shocks are very unlikely to evolve into the RW or SS solutions in astrophysical applications such as failed supernovae, where the radial dynamic range over which the shock propagates is only a few orders of magnitude. Instead, shocks with Mach numbers of order the CQR value can be adequately described by the CQR solution, as shown specifically in Paper I. 

We study the post-shock solution in detail for $n=2.5$. We find that shocks with Mach numbers near a self-similar value have post-shock density, velocity, and pressure profiles that resemble the self-similar solutions (e.g., Fig.\,\ref{fig:fgh}). A striking result is the post-shock flow structure for simulations with Mach numbers that span the gap between the three self-similar solutions: the post-shock flow structure itself continuously varies between the appropriate self-similar values.   Moreover, within a given simulation, the Mach number changes monotonically with time (e.g., Fig.\,{\ref{fig:dmach_time}}) and the post-shock flow structure changes in time as well, effectively bridging the gap between our simulations with different initial Mach numbers. Taking into account both the results of different simulations and the time evolution within a given simulation, we find that shock trajectories can be described by a single function, $\alpha_n(M_{\rm s})$ (i.e., temporal power-law, Eq.\,\ref{eq:ode}, as a function of Mach number), for each value of the density power-law index, $n$, of the ambient medium. 

We find shock trajectories attract toward a stable equilibrium solution, $\alpha_n(M_s)$, for each $n$. The RW and SS solutions are asymptotic limits to the stable equilibrium solution. For $n<2$, shock trajectories migrate away from the SS solution and toward the RW solution. In terms of the evolutionary state at $t\rightarrow\infty$, we refer to the SS solution as an unstable asymptotic limit, while the RW solution is a stable asymptotic limit. For $n>3.5$, we find the roles reverse: the SS (RW) solution is a stable (an unstable) asymptotic limit. Between $2<n<3.5$, the CQR solution is an unstable equilibrium point from which shock trajectories repel
; both RW and SS solutions are stable asymptotic limits. Our linear stability analysis of the CQR solution in Paper II is an analysis of the trajectories near the unstable equilibrium point. The embedded diagram in Fig.\,\ref{fig:mega_alpha} illustrates this interpretation of our analytical and numerical results. 

\cite{1999ApJ...510..379M} presented a method for combining Sedov-Taylor strong shock solutions with accelerating strong shock solutions for exponential atmospheres \citep{sakurai}. This has the attractive feature of fully describing shock propagation through a star.   A similar method may exist for the lower energy shocks considered in this paper (see Eq.\,\ref{eq:ode} and \S\ref{sec:megaalpha}). Our solutions are limited to shocks in point mass gravitational fields and polytropic envelopes, but such conditions are satisfied over many decades in radii in supergiant envelopes. The evolution of a shock through an envelope with a more complicated density profile (not a single power-law) can likely, we suspect, be modeled by an appropriate combination of the numerical solutions presented in this paper. This is an interesting direction for future study because such a solution would be helpful in estimating the ejecta distribution and the rate of accretion onto the central black hole (\S 5.2) in low-energy stellar explosions.


\section*{Acknowledgements}
We thank Paul Duffell, Chris Matzner, and Frank Timmes for useful discussions. This research was funded by the Gordon and Betty Moore Foundation through Grant GBMF5076. ERC acknowledges support from NASA through the Einstein Fellowship Program, Grant PF6-170170. EQ thanks the theoretical astrophysics group and the Moore Distinguished Scholar program at Caltech for their hospitality and support. This work was supported in part by a Simons Investigator Award from the Simons Foundation. 

The software used in this work was in part developed by the DOE NNSA-ASC OASCR Flash Center
at the University of Chicago.

\software{FLASH \citep{2000ApJS..131..273F}}

\appendix
\twocolumngrid
\renewcommand\thefigure{\thesection.\arabic{figure}}    
\setcounter{figure}{0}    
\section{Simulation Parameters and Convergence}
\begin{figure}[]
  \centering
    \includegraphics[width=\columnwidth]{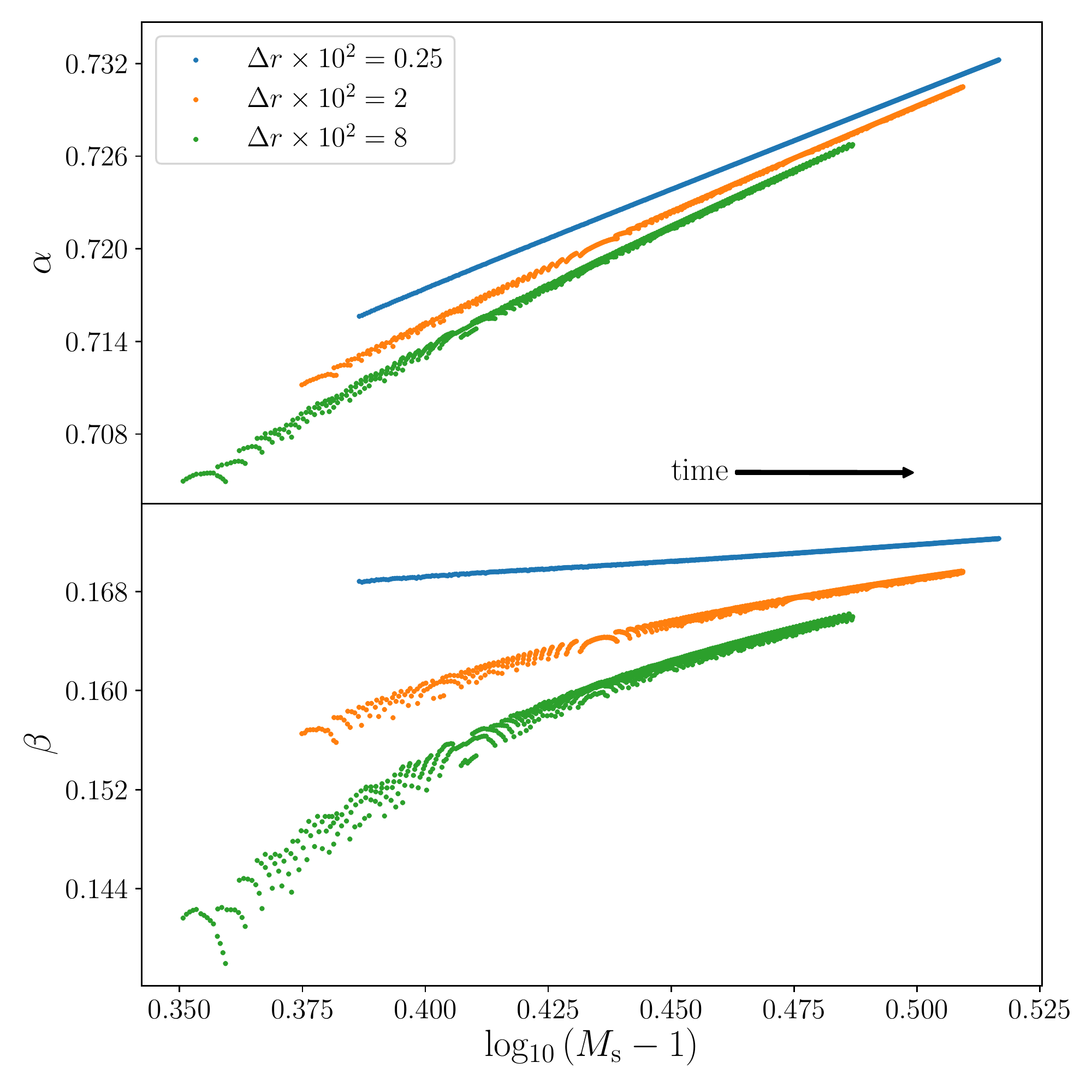}
  \caption{Temporal power-law (Eq.\,\ref{eq:alpha}) and growth rate (Eq.\,\ref{eq:sigma2}) of a shock with $\delta p = 1$ and $n=2.5$ between times $10^2\le t\le 10^3$. Three sets of simulations are shown with different grid resolutions. }  
  \label{fig:convergence}
\end{figure}

The effects from grid resolution are studied for $n=2.5$ and $\delta p=1$. We consider a uniform grid with high, medium, and low resolution: $\Delta r \times 10^2=0.25$, 2, and 8. For reference, all of our simulations in the main text have $\Delta r \times 10^2 \le 4$. The temporal power-laws, $\alpha$, and growth rates, $\beta$, (with respect to the CQR solution) are shown in Fig.\,\ref{fig:convergence} for times between $10^2\le t \le 10^3$. Since the grid is discretized, the prescribed location of the over-pressurized boundary, $r_i=1.5$, almost always resides within a grid cell rather than at a grid cell boundary. This effectively changes the initial explosion energy of the simulation which consequently affects the Mach numbers seen in Fig.\,\ref{fig:convergence}. While this effect is easily correctable, it is not significant for our studies. Fig.\,\ref{fig:convergence} shows the differences in $\alpha$ and $\beta$ between the low and high resolution runs is relatively small. 

The primary benefit of sufficiently high grid resolution is the improved accuracy of the shock location and jump conditions. These were important for our detailed analyses in Paper II. A high resolution is generally not necessary to study the asymptotic trajectories and post-shock distributions of stronger shocks (although, see Appendix B). 

\begin{table}[]
\begin{center}
\setlength{\tabcolsep}{2pt}
\renewcommand{\arraystretch}{0.75} 
\begin{tabular}{ |c|c|rrr| } 

\hline
$n$ & $\Delta r\times10^{2}$  & & $\delta p\ \left[M_s\left(t=10^2\right)\right]$ & \\
\hline
1.5 & 2 & -0.5 [1.01] & 0 [1.08] & 1 [1.21]     \\
 & &  3 [1.42]  & &  \\
 & 4 & 10 [2.00] & 30 [2.89] & 100 [4.73]   \\
 & & 300 [7.53] & $10^3$ [10.78] & $10^4$ [28.99] \\
 & & $10^5$ [77.91] & &  \\
\hline
2.0 & 2 & -0.5 [1.05] & 0.1 [1.25] & 1 [1.62]  \\
 & 4 & 3 [2.43] & 10 [4.49] & 100 [14.60]  \\
 & &  $10^3$ [46.60] & & \\
\hline
2.1 & 2 & -0.5 [1.07] & 0 [1.26] & 1 [1.81]   \\ 
 & & 2 [2.35]  & 3 [2.95] & 4 [3.47]   \\ 
 & &  4.5 [3.67] & 5 [4.01] &  5.5 [4.18] \\
 & & 6 [4.36] &  7 [4.82] & 10 [5.88]  \\
 & 4 & 30 [10.72] & 100 [20.22] & 300 [35.68]  \\
\hline
2.5 & 0.0625 & 0.1 [1.80] & &    \\ 
 & 0.25 & 1 [3.44] & &    \\ 
 & 1 & 0.25 [1.86] &  0.275 [2.06] &  0.6 [2.62]   \\ 
 & 2 & -0.65 [1.05] & -0.55 [1.11] & -0.5 [1.14]  \\
 & & -0.35 [1.26] & -0.25 [1.35] & -0.1 [1.50]  \\
 & & 0 [1.63] & 0.1 [1.76] & 0.125 [1.79]  \\ 
 & & 0.15 [1.83] & 0.175 [1.86] & 0.2 [1.90] \\
 & & 0.27 [2.008] & 0.275 [2.015] & 0.277 [2.019] \\ 
 & &  0.28 [2.023]  & 0.225 [1.94] & 0.25 [1.98]  \\
 & & 0.26 [1.99] & 0.3 [2.06] & 0.4 [2.22]  \\
 & & 0.5 [2.39] & 0.6 [2.56] & 0.7 [2.74] \\
 & & 0.8 [2.92] & 0.9 [3.11] & 1 [3.35]  \\
 & &  1.5 [4.30] & 2 [5.18] & 3 [6.78]  \\
 & & 5 [9.61] & 10 [14.88] & 30 [28.98] \\
 & & 100 [61.56]  & & \\
\hline
2.9 & 1 & -0.65 [1.13] & -0.625 [1.16] & -0.6 [1.19]  \\
 & &  -0.575 [1.22] & -0.55 [1.26] & -0.525 [1.29]   \\
 & & -0.5 [1.33] & -0.475 [1.38] & -0.45 [1.42]  \\
 & & -0.4 [1.52] & & \\
 & 4 & -0.3 [1.77] & -0.2 [2.08] & -0.1 [2.46]   \\
 & & 0 [2.89] & [0.5 5.59] & 1 [8.83]    \\
 & & 3 [19.67] & 10 [43.32] & 100 [175.26] \\
\hline
3.0 & 1 &  -0.7 [1.07] & -0.65 [1.15] & -0.6 [1.23] \\ 
 & & -0.575 [1.28] & -0.55 [1.32] & -0.5 [1.42]   \\
 & & -0.4 [1.68] & -0.25 [2.22] & -0.15 [2.71] \\
 & 4 & 0.1 [4.26] & 1 [11.56] & 3 [25.90]  \\
 & & 10 [61.30] & & \\
\hline
3.3 & 0.5 & -0.775 [1.08] & -0.765 [1.09] & -0.75 [1.12]  \\
& & -0.7 [1.23] & & \\
 & 4 & -0.6 [1.47] & -0.55 [1.68] & -0.5 [1.92] \\
 & &  -0.45 [2.23]  &  -0.4 [2.60] & -0.3 [3.58]  \\
 & & -0.25 [4.16] & -0.2 [4.80] &  -0.1 [6.23] \\
 & & 0.1 [9.50] & 0.2 [11.28] & 0.4 [14.92]   \\
 & & 1 [25.54] & 2 [45.67] & \\
\hline
3.6 & 0.5 & -0.825 [1.04] & -0.81 [1.07] & -0.8 [1.10] \\
 & 4 & -0.75 [1.21] & -0.74 [1.25] & -0.7 [1.39]   \\ 
 & & -0.65 [1.69] & -0.6 [2.10]  &  -0.55 [2.65]   \\
 & &  -0.5 [3.36] & -0.4 [5.39] & -0.3 [8.81]  \\
 & & -0.1 [15.01] & 0.5 [39.03] & \\
 \hline
\end{tabular}
\end{center}
\caption{The polytropic indices, grid resolutions, initial pressure jump and shock Mach number at $t=10^2$ for our simulations.  }
\label{table:1}
\end{table}

\section{Hollow Interiors for $\MakeLowercase{n}>6/(\gamma+1)$}
\label{sec:hollow}

\begin{figure}[ht]
\centering
\includegraphics[width=\columnwidth]{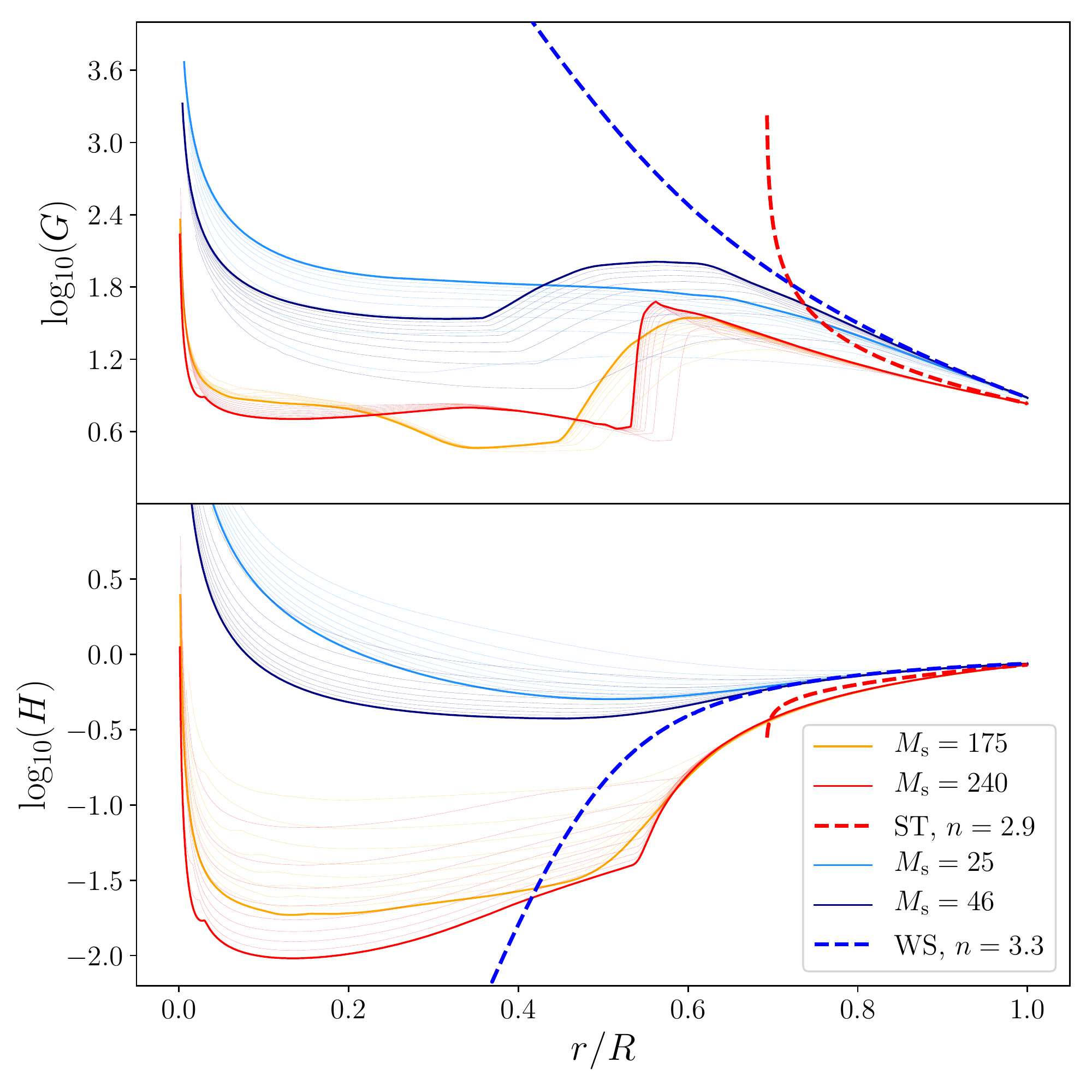}
\caption{Solid lines are the density and pressure distributions behind the shock as defined by Eq.\,(\ref{eq:dimensionless}). Dashed lines are the self-similar solutions for infinitely strong shocks. Red colours are for $n=2.9$ and blue colours are for $n=3.3$. Mach numbers in the legend are measured at $t=10^2$. The grid resolution for the highest Mach number solution (i.e., $M_{\rm s}=240$) is $8\times$ finer than the other simulations, which have the same grid resolution. The Sedov-Taylor solution contains a contact discontinuity with infinite density and zero pressure at $r/R\simeq0.7$. The solution by \cite{ws93} diverges toward $r=0$ and has a sonic point at $r/R\simeq0.608$. }
\label{fig:gh_n2p9}
\end{figure}

In Fig.\,\ref{fig:gh_n2p9}, we compare the numerical post-shock density and pressure distributions with the Sedov-Taylor solution for $n=2.9$ (red colours). The strongest shock in Fig.\,\ref{fig:mega_alpha} is the shock shown here along with an identical simulation with a higher grid resolution. We see that the material near the shock front (i.e., $r/R\gtrsim0.9$) resembles the self-similar solution while the remaining distribution does not. The CD is around $r/R\simeq0.55$ and more identifiable in the density profile. We can see that the density surrounding the CD increases very slowly with time. The pressure and density below the CD continuously decline; although, the density decreases extremely slowly. This figure suggests that the shock generated from our initial conditions is approaching the Sedov-Taylor solution but at an extremely slow pace. It is not surprising then that the temporal power-laws from both simulations, which are nearly identical and slowly increasing, also converge slowly toward the Sedov-Taylor value (see Fig.\,\ref{fig:mega_alpha}). 

Also shown in Fig.\,\ref{fig:gh_n2p9} are the numerical solutions for the post-shock density and pressure for $n=3.3$ (blue colours). We find our numerical solutions are in agreement with the analytic solutions from \cite{ws93} for a larger domain in $r/R$. The temporal power-laws in Fig.\,\ref{fig:mega_alpha} do not show a significant discrepancy with the predicted values from \cite{ws93}. This is because these strong shocks develop a sonic point downstream  at $r_{\rm sp}/R\simeq0.608$. The sonic point causally disconnects the shock evolution from details of the supersonic (relative to the shock) post-shock distribution, regardless of whether the supersonic distribution is accurate or not. In contrast, the sonic point in the Sedov-Taylor solutions for $n=2.9$ and 3 are at the CD. So, the entire post-shock distribution needs to be captured to reproduce the Sedov-Taylor solutions. \cite{2010ApJ...723...10K} find success in resolving the CD using a Lagrangian scheme for their numerical calculations.

\section{Empirical Relations for $\alpha_{\MakeLowercase{n}}(M_{\MakeLowercase{s}} )$}
\label{sec:empiricalrelation}
Our numerical solutions suggest there is a relationship, $\alpha_n(M_{\rm s})$, between the temporal power-law and Mach number of the shock trajectory. The numerical solutions are well approximated by either a sigmoid function, a probability distribution function, or a combination of the two. In Table\,\ref{table:2}, we present a list of empirical relations that approximate the numerical solutions of $\alpha_n(M_{\rm s})$ over the domain of Mach numbers in Fig.\,\ref{fig:mega_alpha}. As discussed in Appendix B, our solutions do not fully reproduce the analytic solutions for $n=2.9$ and 3. It is therefore likely that a more physical best fit function would include replacing the terms 0.925 and 0.958 with $\alpha_{\rm ST}$ in Table\,\ref{table:2}.

\begin{table}[ht]
\begin{center}
\setlength{\tabcolsep}{3.75pt}
\renewcommand{\arraystretch}{0.75} 
\begin{tabular}{ |c|c| } 
\hline
$n$ &  $\alpha_n(M_s)$\\
\hline
1.5 & $\alpha_0 - 0.82 \left(8.50 +  e^{-3x}\right)\inv$ \\
2.0 & $ \left(\alpha_0\inv + e^{-1.48 x^2 - 0.76 x - 2.40} \right)\inv$ \\
2.1 & 
  $\begin{cases}
   \alpha_0 - 0.5\, x^2 e^{2.8 x} ,\  \text{\ \ \ \ \ \ \ \ \ \ \ \ \ \ \ \ \ if}\ x\le0,\\ \nonumber
    \alpha_{\rm ST} -0.036\left(0.67+e^{4.5x} \right)\inv,\ \text{if}\ x\ge0,\nonumber
  \end{cases}$
\\
2.5 & 
$
  \begin{cases}
   \alpha_0 +0.094\, x e^{2.26 x} ,\ \text{\ \ \ \ \ \ \ \ \ \ \ \ \ \ if}\ x\le0,\\ \nonumber
    \alpha_{\rm ST} -0.77 \left(4.77+e^{3.36 x}\right)\inv,\ \text{if}\ x\ge0,\nonumber
  \end{cases}
$
 \\
2.9 & 
$
\begin{cases}
   \alpha_0 + 0.043\, x e^{2.8 x} ,\  \, \text{\ \ \ \ \ \ \ \ \ \ \ \ \ \ if}\ x\le0,\\ \nonumber
    0.925 - 4.0\left(14.4+e^{2.9 x}\right)\inv,\ \text{if}\ x\ge0,\nonumber
  \end{cases}
$
 \\
3.0 & 
$
  \begin{cases}
   \alpha_0 + 0.0084\, \sqrt{|x|} e^{0.8 x} ,\ \text{\ \ \ \ \ \ \ \  if}\ x\le0,\\ \nonumber
    0.958 - 6.2\left(20.2+e^{2.8 x}\right)\inv,\ \text{if}\ x\ge0,\nonumber
  \end{cases}
$ 
 \\
3.3 & $\alpha_{\rm WS} - 19.7\, e^{0.055x}\left(50.7 + e^{2.6x}\right)\inv$ \\
3.6 & $\alpha_{\rm WS} - 1.89 \, e^{0.027 x}  \left(3.85 + e^{2.6x}\right)\inv $ \\
\hline
\end{tabular}
\end{center}
\caption{Empirical relations that approximate the numerical solutions for $\alpha_n(M_{\rm s})$ in Fig.\,\ref{fig:mega_alpha}. The shock Mach number is defined as $x={\rm log}_{10}(M_{\rm s}/M_0)$ where $M_0$ is the Mach number of the CQR solution. For $n\le2$ or $n\ge3.5$, $M_0=1$ since the CQR solution does not exist. The constants, $\alpha_i$, are the temporal power-laws for the RW or CQR solution, $\alpha_0=2/3$, the Sedov-Taylor solution, $\alpha_{\rm ST}=2/(5-n)$, and \citeauthor{ws93} solution, $\alpha_{\rm WS} \simeq 1.0494$ ($n=3.3$) or $1.136$ ($n=3.6$).
}
\label{table:2}
\end{table}


\begin{thebibliography}{}
\expandafter\ifx\csname natexlab\endcsname\relax\def\natexlab#1{#1}\fi

\bibitem[{{Bondi}(1952)}]{1952MNRAS.112..195B}
{Bondi}, H. 1952, \mnras, 112, 195

\bibitem[{Book(1994)}]{DLBook1994}
Book, D.~L. 1994, Shock Waves, 4, 1

\bibitem[{{Chevalier}(1976)}]{1976ApJ...207..872C}
{Chevalier}, R.~A. 1976, \apj, 207, 872

\bibitem[{{Coughlin} {et~al.}(2018{\natexlab{a}}){Coughlin}, {Quataert},
  {Fern{\'a}ndez}, \& {Kasen}}]{2018MNRAS.477.1225C}
{Coughlin}, E.~R., {Quataert}, E., {Fern{\'a}ndez}, R., \& {Kasen}, D.
  2018{\natexlab{a}}, \mnras, 477, 1225

\bibitem[{{Coughlin} {et~al.}(2018{\natexlab{b}}){Coughlin}, {Quataert}, \&
  {Ro}}]{paper1}
{Coughlin}, E.~R., {Quataert}, E., \& {Ro}, S. 2018{\natexlab{b}}, \apj, 863,
  158

\bibitem[{{Coughlin} {et~al.}(2019){Coughlin}, {Ro}, \& {Quataert}}]{paper2}
{Coughlin}, E.~R., {Ro}, S., \& {Quataert}, E. 2019, arXiv e-prints,
  arXiv:1901.04487

\bibitem[{Courant \& Friedrichs(1999)}]{courant1999supersonic}
Courant, R., \& Friedrichs, K. 1999, Supersonic Flow and Shock Waves, Applied
  Mathematical Sciences (Springer New York)

\bibitem[{{Ertl} {et~al.}(2016){Ertl}, {Janka}, {Woosley}, {Sukhbold}, \&
  {Ugliano}}]{ertl16}
{Ertl}, T., {Janka}, H.-T., {Woosley}, S.~E., {Sukhbold}, T., \& {Ugliano}, M.
  2016, \apj, 818, 124

\bibitem[{{Fern{\'a}ndez} {et~al.}(2018){Fern{\'a}ndez}, {Quataert},
  {Kashiyama}, \& {Coughlin}}]{2018MNRAS.476.2366F}
{Fern{\'a}ndez}, R., {Quataert}, E., {Kashiyama}, K., \& {Coughlin}, E.~R.
  2018, \mnras, 476, 2366

\bibitem[{{Fryxell} {et~al.}(2000){Fryxell}, {Olson}, {Ricker}, {Timmes},
  {Zingale}, {Lamb}, {MacNeice}, {Rosner}, {Truran}, \&
  {Tufo}}]{2000ApJS..131..273F}
{Fryxell}, B., {Olson}, K., {Ricker}, P., {et~al.} 2000, \apjs, 131, 273

\bibitem[{{Goodman}(1990)}]{1990ApJ...358..214G}
{Goodman}, J. 1990, \apj, 358, 214

\bibitem[{{Horiuchi} {et~al.}(2011){Horiuchi}, {Beacom}, {Kochanek}, {Prieto},
  {Stanek}, \& {Thompson}}]{2011ApJ...738..154H}
{Horiuchi}, S., {Beacom}, J.~F., {Kochanek}, C.~S., {et~al.} 2011, \apj, 738,
  154

\bibitem[{Hugoniot(1887)}]{ecole1887journal}
Hugoniot, P.~H. 1887, Sur La Propagation Du Mouvement Dans Les Corps et
  Specialement Dans Les Gaz Parfaits (Ecole polytechnique)

\bibitem[{{Krumholz} {et~al.}(2007){Krumholz}, {Klein}, {McKee}, \&
  {Bolstad}}]{2007ApJ...667..626K}
{Krumholz}, M.~R., {Klein}, R.~I., {McKee}, C.~F., \& {Bolstad}, J. 2007, \apj,
  667, 626

\bibitem[{{Kushnir} \& {Waxman}(2010)}]{2010ApJ...723...10K}
{Kushnir}, D., \& {Waxman}, E. 2010, \apj, 723, 10

\bibitem[{{Lovegrove} \& {Woosley}(2013)}]{2013ApJ...769..109L}
{Lovegrove}, E., \& {Woosley}, S.~E. 2013, \apj, 769, 109

\bibitem[{{Matzner} \& {McKee}(1999)}]{1999ApJ...510..379M}
{Matzner}, C.~D., \& {McKee}, C.~F. 1999, \apj, 510, 379

\bibitem[{Nadyozhin(1980)}]{Nadyozhin1980}
Nadyozhin, D.~K. 1980, Astrophysics and Space Science, 69, 115

\bibitem[{O'Connor \& Ott(2011)}]{O_Connor_2011}
O'Connor, E., \& Ott, C.~D. 2011, The Astrophysical Journal,
  730, 70

\bibitem[{Rankine(1870)}]{Rankine01011870}
Rankine, W. J.~M. 1870, Philosophical Transactions of the Royal Society of
  London, 160, 277

\bibitem[{Sakurai(1960)}]{sakurai}
Sakurai, A. 1960, Communications on Pure and Applied Mathematics, 13, 353

\bibitem[{Sedov(1946)}]{sedov1946}
Sedov, L.~I. 1946, Prikl. Mat. Mekh, 10, 241

\bibitem[{{Taylor}(1950)}]{1950RSPSA.201..159T}
{Taylor}, G. 1950, Proceedings of the Royal Society of London Series A, 201,
  159

\bibitem[{von Neumann(1941)}]{vonNeumann1941}
von Neumann, J. 1941, The point source solution, {U.S. Government} Document
  {AM-9}, National Defense Research Council, Division B, Washington, DC, USA

\bibitem[{Waxman \& Shvarts(1993)}]{ws93}
Waxman, E., \& Shvarts, D. 1993, Physics of Fluids A: Fluid Dynamics, 5, 1035

\end{thebibliography}

\end{document}